\documentclass{aa}

\usepackage{amssymb}

\usepackage{graphicx}

\usepackage{longtable,tabularx}

\usepackage{natbib}
\bibpunct{(}{)}{;}{a}{}{,}

\usepackage[bookmarks=true]{hyperref}

\begin{document}

\title{3D dynamical evolution of the interstellar gas\\ in the Gould Belt}
\titlerunning{3D dynamical evolution of the Gould Belt}

\author{Christophe A. Perrot \and Isabelle A. Grenier}
\authorrunning{Perrot,C.A. \and Grenier I.A.}

\offprints{C.~Perrot, \email{cperrot@discovery.saclay.cea.fr}}
\institute{Universit\'e Paris VII \& CEA/Saclay, Service
d'Astrophysique, 91191 Gif-sur-Yvette, France \\
            \email{cperrot@discovery.saclay.cea.fr, isabelle.grenier@cea.fr}}
\date{received xxx / accepted yyy}
\abstract{The dynamical evolution of the Gould Belt has been
modelled in 3D and confronted to the spatial and velocity
distributions of all HI and H$_{2}$ clouds found within a few
hundred parsecs from the Sun and to the \textit{Hipparcos}
distances of the nearby OB associations. The model describes the
expansion of a shock wave that sweeps momentum from the ambient
medium. It includes the effects of the Galactic differential
rotation and its gravitational torque, as well as interstellar
density gradients within and away from the Galactic plane,
possible fragmentation and drag forces in the late stages, and an
initial rotation of the system. The evolved Belt geometry and
velocity field have been fitted to the directions and velocities
of the nearby clouds using a maximum-likelihood test. In order to
do so, local clouds have been systematically searched for in the
available HI and CO surveys. The likelihood function also included
a distance estimate for a subset of well-known clouds.

The best fit to the data yields values for the current Belt
semi-axes of $(373 \pm 5)$ pc and $(233 \pm 5)$ pc, and an
inclination of $17.2\degr \pm 0.5\degr$. These characteristics are
consistent with earlier results, but a different Belt orientation
has been found because of the presence of new molecular clouds and
the revised distance information: the Belt centre currently lies
$(104 \pm 4)$ pc away from the Sun, towards the Galactic longitude
$l_{centre} = 180.4\degr \pm 2.0\degr$, and the ascending node
longitude is $l_{\Omega} = 296.1\degr \pm 2.0\degr$. The Belt
characteristics are independent of an initial rotation. The
present Belt rim is found to coincide with most of the nearby OB
associations and H$_{2}$ clouds, but the Belt expansion bears
little relation to the average association velocities and the
younger ones are surprisingly found farther out from the Belt
center. An initial kinetic energy of $(1.0 \pm 0.1)$ 10$^{45}$ J
and an expansion age of $(26.4 \pm 0.4)$ Myr are required, in good
agreement with earlier 2D estimates. The factor of 2 discrepancy
that exists between the dynamical Belt age and that derived from
photometric stellar ages could not be solved by adding a vertical
dimension in the expansion, nor by adding drag forces and
fragmentation, nor by introducing an initial rotation. Allowing
the Belt to cross the Galactic disc before reaching its present
position would require a longer age of 52 Myr, but the very poor
fit to the data does not support this possibility.
}
\maketitle
\section{Introduction}
The Gould Belt is a nearby starburst region where many stars have
formed over 30 to 40 million years in a surprisingly flat and
inclined disc. New facets of its activity have recently emerged at
high energy with the discovery of a population of $\gamma$-ray
sources associated with it
\citep{grenier_2000_art,gehrels_2000_art}. No clear picture has
emerged yet as to the nature of these objects. Neutron star
activity in various forms appears as a promising prospect, in
particular $\gamma$-ray emission from million-year old pulsars. As
supernova relics, these sources have drawn attention to the
enhanced supernova rate inside the Gould Belt, which has been
estimated to be 3 to 5 times higher that in the local Galactic
disc. In other words 20 to 27 supernovae have occurred in the Belt
per million year over the past few million years
\citep{grenier_2000_art}.

\citet{herschel_1847_book} first pointed out that the distribution
of bright stars was asymmetric about the Galactic plane. The
geometry of this young structure (most of the stars are less than
30--60 million years old) was studied in
\citeyear{gould_1874_proc} by \citeauthor{gould_1874_proc} who
determined its inclination and the direction of its poles. We
refer the reader to \citet{poppel_1997_art} for an extensive
review of Gould's Belt system and its relation to the local
interstellar medium. Besides its peculiar geometry, this system of
early-type stars is known to expand, and to rotate in the same
direction as the galactic rotation. The Gould Belt also contains
interstellar clouds and \citet{lindblad_1967_art} gave strong
evidence for the Belt relation to a locally expanding HI ring.
Famous H$_{2}$ complexes, such as Orion and Ophiuchus, are often
mentioned in relation to the Belt and the fact that nearby dark
clouds participate to the Belt expansion was recognized by
\citet{taylor_1987_art}. Since then, the molecular components of
the local interstellar medium have been thoroughly mapped in the
CO surveys, but no correlation study has been undertaken.

\citet{comeron_1994a_art} concluded that 40\% to 50\% of the young
massive stars (with spectral types earlier than B8) that lie
within 450 pc from the Sun belong to the Belt. The new
\textit{Hipparcos} data bring this fraction to 60\% for stars
within 600 pc and a Belt age of 30 to 60 Myr
\citep{torra_2000a_art}. Whereas the massive-star content of the
Belt has been extensively studied, less is known about its
low-mass star production. Young (30--80 Myr old) Lithium-rich
solar-mass stars with active coronae show up as X-ray sources and
nicely trace the Belt in the sky \citep{guillout_1998b_art}. Even
though the X-ray horizon is limited  by interstellar absorption to
100 or 150 pc for these faint sources, their space distribution
indicates that stellar formation is not only active along the Belt
rim, but also inward, over a significant, yet poorly constrained,
radial extent.

Preserving the structural coherence (flatness and inclination) of
the Belt stellar system over a long time span is challenging,
especially over a large fraction of the vertical oscillation and
expansion timescales. \citet{comeron_1999_art} has explored the
stellar vertical motions and found that the rotating stellar disc
was initially tilted. Its rotation axis was not perpendicular to
the Galactic plane as would be expected from the disruption of a
giant rotating molecular cloud. According to
\citet{lindblad_1997_proc}, the dissolution of an unbound rotating
system of stars, possibly born 30 Myr ago in a spiral arm, may
reproduce the observations, and the rotation may explain the
persistence of a flat disc.

The Belt flatness and its tilt, some $20\degr$ to the Galactic
plane, bear important information on the Belt origin, but remain
very difficult to interpret. Various scenarii involve the oblique
impact of a high-velocity HI cloud on the Galactic disc
\citep{comeron_1992_art,comeron_1994b_art} or a cascade of several
supernova explosions (see \citet{poppel_1997_art} for a review).
The measured Oort constants of the stellar field are consistent
with a cloud impact about 50 Myr ago \citep{comeron_1994b_art}. On
the other hand, an explosive event would explain the kinematics of
the cold neutral medium at $|b| \geq 10\degr$
\citep{poppel_2000_art}. In \citeyear{olano_1982_art},
\citeauthor{olano_1982_art} studied the expansion, within the
Galactic plane, of an initially circular shock wave. The 30 Myr
expansion in this 2D model was constrained by the present-day
longitude-velocity distribution of the HI gas in the Lindblad
ring. Following the expansion of a superbubble, with or without
ambient interstellar pressure, \citet{moreno_1999_art} compared
the emerging stellar orbits to the velocity field of nearby
massive stars. They concluded that the Pleiades group, and
possibly the Sco-Cen association, significantly deviate from this
single expansion model. More recently, \citet{olano_2001_art}
presented a 3D model for the local system of gas and stars
associated with the Sirius supercluster, the Gould Belt, and the
Local Arm. The latter two subsystems would have formed in the
braking of a supercloud in a spiral arm while the stars of older
generations, as in the Sirius supercluster, would move on with
their initial velocity field. The supercloud angular momentum
being concentrated at large radii, the inner regions would
collapse into a flattened disc, precursor of the Gould Belt,
whereas the ejection of the outer parts into a super-ring would
form a precursor of the Local Arm.

To improve on Olano's model and to study the Belt relation to the
new molecular clouds and its position with respect to the local
clouds and OB associations, we model in this work the 3D dynamical
evolution of an inclined cylindrical shock wave, expanding in the
interstellar medium in the Galactic plane and at higher altitudes
(\textit{c.f.} section \ref{sec:model}). Using a
maximum-likelihood approach described in section
\ref{ssec:likelihood}, we confront the characteristics of the gas
shell in the longitude-latitude-velocity phase space to the
observed positions and velocities of all nearby HI and H$_{2}$
clouds that have been found in extensive HI and CO surveys (see
section \ref{sec:cloud:selection}). We then discuss the best-fit
evolution and current geometry of the Belt and its relation to the
stars in section \ref{sec:results}.
\section{Evolution scenario and analysis}
\label{sec:model}
\subsection{Dynamical evolution}
The model does not attempt to explain the origin of the outburst,
but describes the lateral expansion of an inclined, cylindrical
shock wave that sweeps momentum from the ambient interstellar
medium. As a consequence of the Galactic differential rotation,
the circular section of the Belt rapidly evolves into an
elliptical one. Additionally, the combined actions of the Galactic
gravitational potential and of the interstellar density gradients
further warp the Belt. The Belt ring was split into 60 elementary
sections. Their contiguous surfaces delineated the Belt rim
without any holes at each time step and allowed it to bend and
warp easily. The motion of each Belt element was simulated, taking
into account the galactic gravitational forces, and the momentum
variation induced by the swept-up gas. The integration was done
using a Runge-Kutta algorithm with an adaptative timestep, so that
the distance step for the fastest gas element is less than 1 pc.

The initial Belt geometry is described by a limited set of free
parameters: its inclination with respect to the Galactic plane,
$\varphi_{0}$; the longitude of its ascending node,
$l_{\Omega}^{0}$; the constant cylinder height, $H$; the longitude
and distance from the Sun of its centre, $d_{centre}^{0}$ and
$l_{centre}^{0}$; its age, $\tau$; the initial mass and velocity
of the ejecta, $M_{0}$ and $v_{0}$. The expelled mass was
uniformly distributed along the Belt. An initial radius of 20 pc,
typical of OB associations, was adopted, but it does not
significantly influence the subsequent evolution. The Belt centre
was assumed to be initially located in the Galactic plane, at $z =
0$. An elliptical fit to the position of the sections allows to
describe the Belt geometry at any stage in terms of its semi-major
axis, $a$, and semi-minor axis, $b$.
\subsection{Interstellar density gradients}
\label{ssec:ISMgrad}
The local HI and H$_{2}$ gas density into which the shock wave
expands is described by a combination of three Gaussian and one
exponential functions \citep{dickey_1990_art,dame_1987_art}:
\begin{center}
\begin{eqnarray}
\rho(z) =
N_{1,\,HI}\;exp(-\frac{z^{2}}{2{\sigma_{1}^2}})+N_{2,\,HI}\;exp(-\frac{z^{2}}{2{\sigma_{2}}^2})\\
+N_{3,\,HI}\;exp(-\frac{|z|}{\sigma_{3}})+N_{H_{2}}\;exp(-\frac{z^{2}}{2{\sigma_{H_{2}}}^2})\nonumber
\end{eqnarray}
\end{center}
with $N_{1,\,HI} = 0.395$ cm$^{-3}$, $\sigma_{1} = 90.03$ pc,
$N_{2,\,HI} = 0.107$ cm$^{-3}$, $\sigma_{2} = 225.17$ pc,
$N_{3,\,HI} = 0.064$ cm$^{-3}$, $\sigma_{3} = 403.0$ pc,
$N_{H_{2}} = 0.2$ cm$^{-3}$, and $\sigma_{H_{2}} = 74.0$ pc. A
radial, galactocentric density gradient was also included,
dropping linearly by a factor of 2 over 1 kpc. It had no impact on
the results, so the sole $z$ dependence on the interstellar
density has been retained hereinafter. In order to compute the
interstellar gas momentum, we adopted the IAU recommended values
for the Oort's constants $A_{c} = 14.5$ km s$^{-1}$ kpc$^{-1}$ and
$B_{c} = -12.5$ km s$^{-1}$ kpc$^{-1}$, the Solar galactocentric
radius $R_{c} = 8.5$ kpc, and its orbital velocity $v_{c} = 220$
km s$^{-1}$. No internal pressure from supernova or stellar winds
was added, nor any external pressure from the interstellar medium.
All calculations were done in a Galactic cartesian inertial frame,
centred on the Galactic centre, the x-axis pointing from the Sun
to the Galactic centre at $t = 0$, the y-axis pointing in the
direction of Galactic rotation, and the z-axis pointing to the
North Galactic pole.
\subsection{The Galactic gravitational potential}
\label{ssec:Fdisc}
The torque locally induced by the gravitational potential of the
Galactic disc was calculated as a function of altitude $z$ using
the local stellar mass density distribution, with a volume density
$\rho_{\star} = 7.6$ $10^{-2}$ M$_{\odot}$ pc$^{-3}$ at $z = 0$
\citep{creze_1998_art} and an exponential scale height $z_{\star}
= 260 \pm 60$ pc \citep{ojha_1996_art}. Assuming that the Galactic
disc can be locally described by an infinite plane, the
corresponding vertical acceleration is given by:
\begin{center}
\begin{eqnarray}
\overrightarrow{a_{z}} & = &
-4\pi\mathcal{G}\rho_{\star}z_{\star}\left[1-\exp\left(-\frac{|z|}{z_{\star}}\right)\right]\frac{\overrightarrow{z}}{|z|}
\end{eqnarray}
\end{center}
$\mathcal{G}$ denoting Newton's gravitational constant.
\subsection{Maximum-likelihood analysis}
\label{ssec:likelihood}
\begin{table}
\caption{Longitude, latitude and velocity ranges for cloud
selection}
\begin{tabularx}{1.0\linewidth}{rrcrr} \hline
 $l_{\mathrm{min}}$ & $l_{\mathrm{max}}$ & $b$ & $v_{\mathrm{min}}$ & $v_{\mathrm{max}}$ \\
 ($\degr$) & ($\degr$) & ($\degr$) & (km s$^{-1}$) & (km s$^{-1}$) \\ \hline
 0    & 10    & $b > 8$             & -8 & 12  \\
 10   & 85    & $|b| > 8$           & -8 & 12  \\
 85   & 100   & $b > 8$             & -8 & 12  \\
 100  & 110   & $b > 5$             & -8 & 12  \\
 110  & 155   & $|b| > 5$           & -8 & 12  \\
 155  & 165   & $b < -12$           &  0 & 12  \\
 155  & 170   & $-12 < b < -5$      & -8 &  0  \\
 170  & 180   & $b > 8$             & -8 & 12  \\
 -180 & -175  & $b > 8$             & -8 & 12  \\
 -175 & -160  & $|b| > 8$           & -8 & 12  \\
 -160 & -100  & $|b| > 8$           & -8 & 9.5 \\
 -100 & -80   & $|b| > 8$           & -8 & 12  \\
 -80  & -65   & $|b| > 5$           & -8 & 12  \\
 -65  & -55   & $b > 5$             & -8 & 12  \\
 -65  & -55   & $-12 < b < -5$      & -8 & 12  \\
 -55  & -40   & $|b| > 5$           & -8 & 12  \\
 -40  & -3    & $|b| > 8$           & -8 & 12  \\
 -3   & 0     & $b > 8$             & -8 & 12  \\ \hline
\end{tabularx}
\label{tab:cuts}
\end{table}
\begin{figure*}
\includegraphics[width=\linewidth]{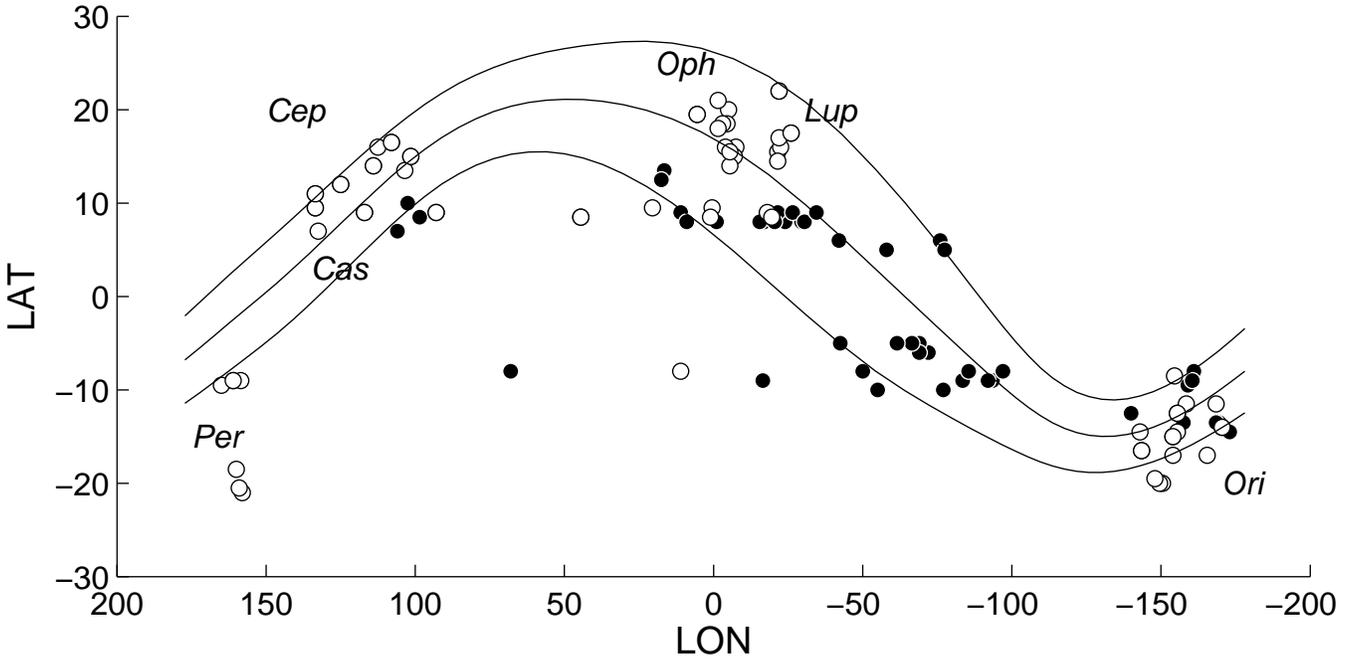}
\caption{Galactic longitude and latitude distribution of the
nearby HI (black dots) and H$_{2}$ (white dots) clouds. The dots
mark the direction of peak emission. Clouds at $|b| < 5\degr$ or
$8\degr$ have not been selected. The curves follow the trace of
the Belt across the sky as obtained from the best fit. The upper
and lower lines mark the corresponding edges of the 60 pc thick
disc of the Belt.} \label{fig:lonlat}
\end{figure*}
Varying the initial expansion parameters, the geometry and
velocity field of the evolved Belt (today) have been compared with
the space and velocity distributions of the nearby HI and H$_{2}$
clouds. Because the direction of these clouds is known, but rarely
their distance, we developed a specific maximum-likelihood test to
compare their direction and radial velocity with those of the
modelled Belt sections. The likelihood expression is based on the
probability $\Psi_{ij}$ for a cloud $i$ to be seen at relative
distance $D_{ij}$ from an elementary Belt section $j$, and at a
relative radial velocity $v_{ij}$ with respect to that of the Belt
section (\textit{c.f.} Equation \ref{equ:psi}). Gaussian
probability density functions have been adopted in relative
distance and relative radial velocity with standard deviations
$\sigma_{D} = 20$ pc and $\sigma_{v} = 1$ km s$^{-1}$,
respectively. So, $\Psi_{ij}$ is given by
\begin{center}
\begin{eqnarray}
\Psi_{ij} & = &
\frac{g_{s}}{4\;\pi^{2}\;\sigma_{D}^{3}\;\sigma_{v}}\;e^{-\frac{D_{ij}^{2}}{2\;\sigma_{D}^{2}}}\;e^{-\frac{v_{ij}^{2}}{2\;\sigma_{v}^{2}}}
\label{equ:psi}
\end{eqnarray}
\end{center}
where $1/g_{s}$ is the Belt perimeter.

When the cloud distance to the Sun, $D_{i} \pm \Delta D_{i}$, is
available (\textit{c.f.} Tables \ref{tab:clumpsHI} and
\ref{tab:clumpsCO}), the cloud likelihood is obtained by
integrating over the 60 Belt sections:
\begin{center}
\begin{eqnarray}
\mathcal{L}_{i} & = &
\int_{Belt}\;\Psi_{ij}\;D_{i}^{2}\;\Delta\Omega_{i}\;\Delta
D_{i}\;\Delta v_{i}\;ds_{j}
\end{eqnarray}
\end{center}
$\Delta \Omega_{i} = \Delta l_{i}\;\Delta b_{i}\;\cos b_{i}$
represents the 1 $\sigma$ confidence solid angle around the
direction ($l_{i}, b_{i}$) of peak emission in the cloud, in
Galactic coordinates. $\Delta v_{i}$ notes the 1 $\sigma$ error on
the observed cloud centroid velocity.

For a cloud with no distance estimate, we performed a
Gau\ss-Legendre integration of the spatial probability density
function along the cloud $(l_{i},b_{i})$ direction:
\begin{center}
\begin{eqnarray}
\nonumber
\mathcal{L}_{i} & = &
\int_{Belt}\left[\frac{g_{s}}{4\;\pi^{2}\;\sigma_{D}^{3}\;\sigma_{v}}\;
e^{-\frac{v_{ij}^{2}}{2\;\sigma_{v}^{2}}}\;\Delta\Omega_{i}\;
\Delta v_{i}\;\;\;\;\cdots\right.\\
\cdots & & \left.\int_{\delta = 0}^{+\infty}
\delta^{2}\;e^{-\frac{(\delta\cos l_{i}\cos
b_{i}-x_{j})^{2}}{2\;\sigma_{D}^{2}}}\;\;\;\;\cdots\right.\\
\nonumber
\cdots & & \left.e^{-\frac{(\delta\sin l_{i}\cos
b_{i}-y_{j})^{2}}{2\;\sigma_{D}^{2}}}\;e^{-\frac{(\delta\sin
b_{i}-z_{j})^{2}}{2\;\sigma_{D}^{2}}}\;d\delta\right]\;ds_{j}
\end{eqnarray}
\end{center}
$x_{j}$, $y_{j}$, and $z_{j}$ note the cartesian coordinates of
the centre of a Belt section.

The likelihood $\mathcal{L}(\varphi_{0}, l_{\Omega}^{0}, H,
d_{c}^{0}, l_{c}^{0}, \tau, M_{0}, v_{0})$ of a model is obtained
by the product of individual (independent) cloud probabilities
$\mathcal{L}_{i}$. The maximum likelihood is found using the
\textit{downhill simplex method} developed by
\citet{nelder_1965_art}. The iteration stops when the difference
between the function values at each point of the simplex and the
current minimum are less than $10^{-3}$. This procedure has been
successfully tested using simulated clouds randomly generated from
a known Belt.
\section{Cloud selection}
\label{sec:cloud:selection}
As noted by several authors, the Gould Belt has left its imprint
on the local gas dynamics as seen in the HI data. The local
molecular clouds, as mapped in CO, appear to follow a
longitude-velocity pattern similar to that of the atomic gas in
the Lindblad ring \citep{dame_1987_art}. To include all nearby
atomic and molecular clouds, we used the extensive CO survey of
the Milky Way at $|b| < 25\degr$ by \citet{dame_1987_art} and two
complementary HI surveys
\citep{hartmann_1997_book,strong_1982_art} to cover the whole sky.
Clouds were systematically searched for and isolated using the
powerful \textsc{clumpfind} tool developed by
\citet{williams_1994_art}. This algorithm searches for coherent
peaks of emission based on closed contours in a
longitude-latitude-velocity data cube, from the highest intensity
to the lowest. The algorithm yields the position of peak emission
and centroid radial velocity of the clouds.

The CO data cube was processed in the $[-15, +17]$ km s$^{-1}$
velocity range characteristic of the local medium. The temperature
increment for the \textsc{clumpfind} to follow down the intensity
levels was set to $\Delta T = 1$ K. The same procedure was applied
to the HI data cubes for velocities $|v| < 20$ km s$^{-1}$ and
latitudes restricted to $|b| > 3\degr$ to leave out extensions to
medium latitudes of background clouds in the Galactic plane.
Temperature steps of $\Delta T = 0.07$ K and $\Delta T = 2$ K have
been used for the \citet{hartmann_1997_book} and the
\citet{strong_1982_art} surveys, respectively.

\begin{figure}
\includegraphics[width=\linewidth]{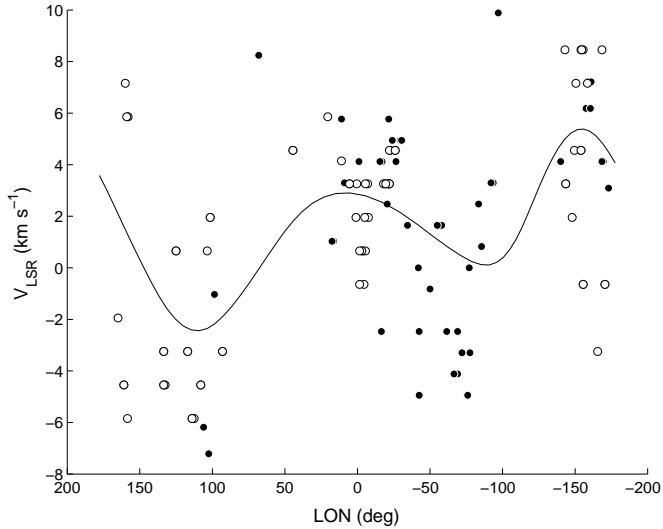}
\caption{longitude-velocity distribution of the nearby HI (black
dots) and H$_{2}$ (white dots) clouds and the best fit to the Belt
rim kinematics. Projected velocities are given in the Local
Standard of Rest.}\label{fig:lonvit}
\end{figure}

Table \ref{tab:cuts} summarizes the selection criteria used in
longitude, latitude and velocity to avoid contamination from
background clouds in the Galactic disc or in the Local Arm. These
cuts were kept minimal to limit biases and to avoid an a priori
selection of clouds obviously linked to the Belt while retaining
the intrinsic spatial and velocity dispersion of the clouds in our
neighbourhood. The lists of selected HI and H$_{2}$ clouds and
their characteristics are given in Tables \ref{tab:clumpsHI} and
\ref{tab:clumpsCO}, respectively. The (l,b) directions and
velocities of the selected clouds are displayed in Figures
\ref{fig:lonlat} and \ref{fig:lonvit}.
\begin{figure}
\includegraphics[width=\linewidth]{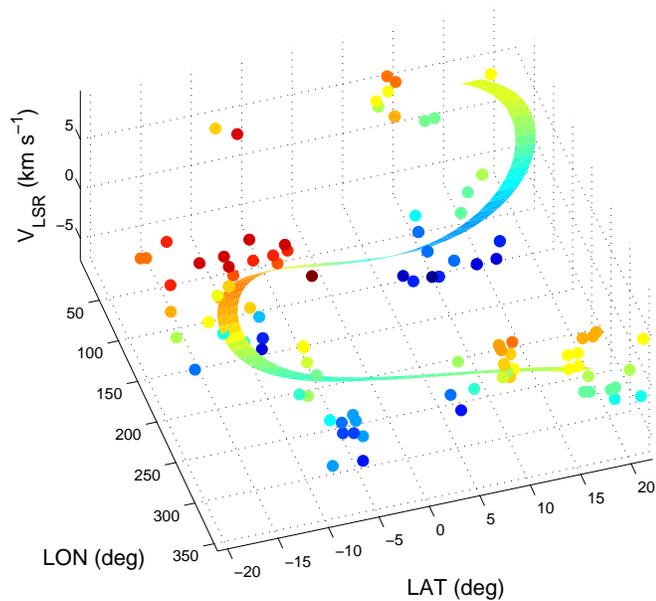}
\caption{longitude, latitude, and velocity distribution of the
nearby HI (black dots) and H$_{2}$ (white dots) clouds and the
best fit to the Belt rim kinematics. Projected velocities are
given in the Local Standard of Rest and are colour-coded from
black to white from negative to positive
values}\label{fig:lonlatvit}
\end{figure}
\section{Results and discussion}
\label{sec:results}
The initial and current characteristics of the Belt expansion that
best fit the present-day cloud data are given in Table
\ref{tab:fitparam}. The resulting geometry and dynamics appear to
be independent of the presence or not in the fit of the Taurus
clouds which, since they lie near the centre, do not participate
to the expansion of the Belt rim . The quoted 1$\sigma$ errors are
purely statistical and have been determined from the
log-likelihood ratios and the information matrix
\citep{strong_1985_art}.

\begin{figure*}
\centerline{\includegraphics[width=0.45\linewidth]{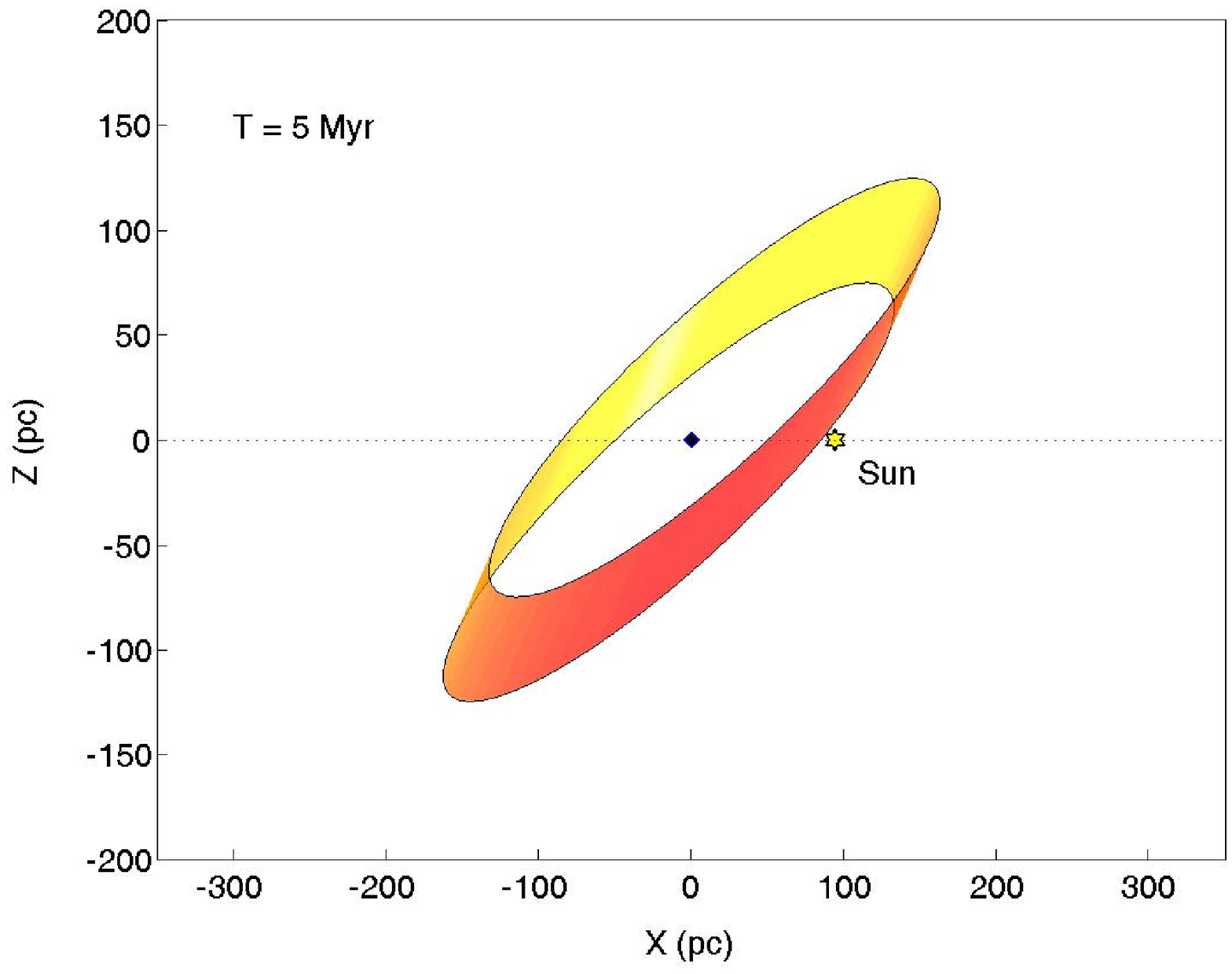}
\includegraphics[width=0.45\linewidth]{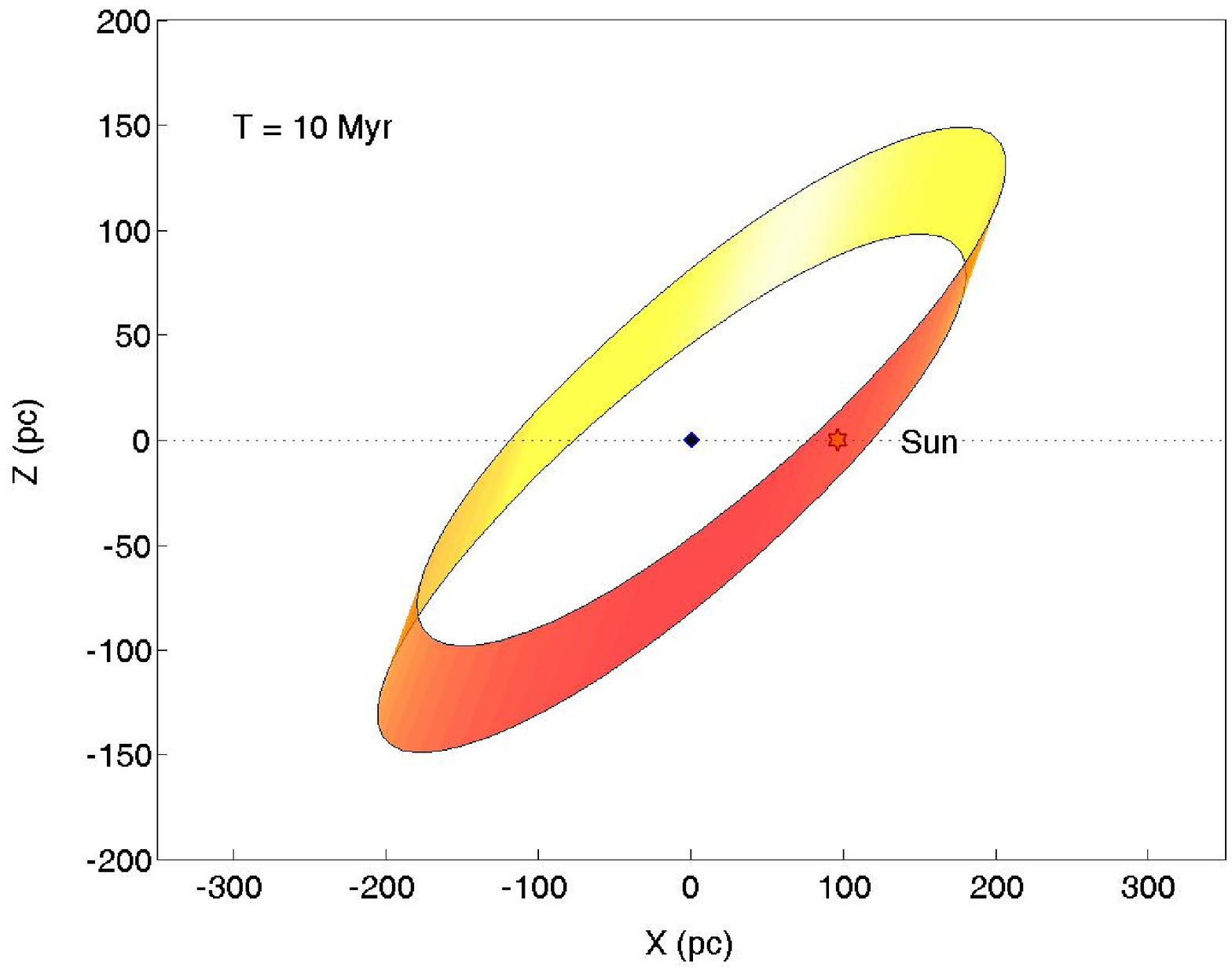}}
\centerline{\includegraphics[width=0.45\linewidth]{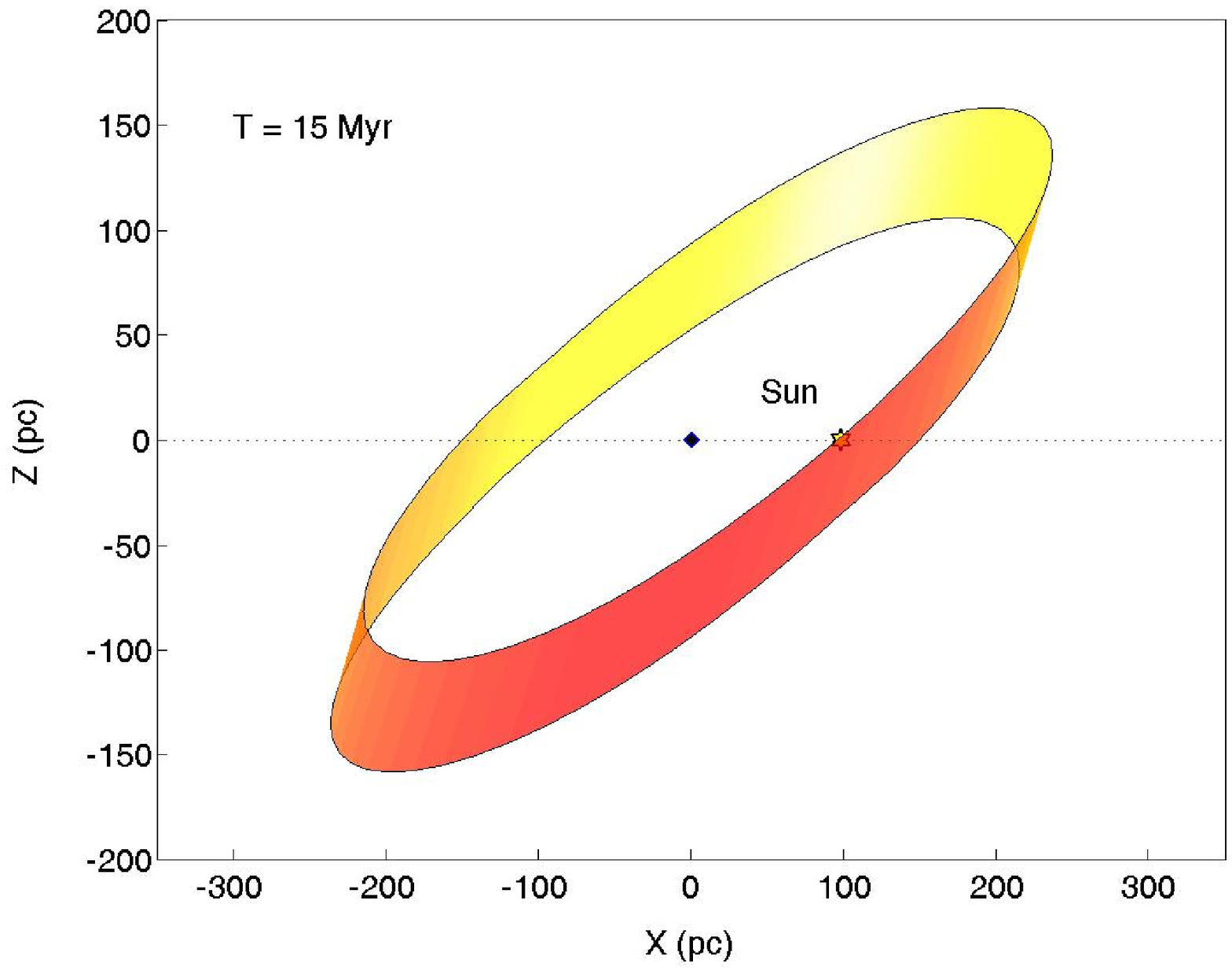}
\includegraphics[width=0.45\linewidth]{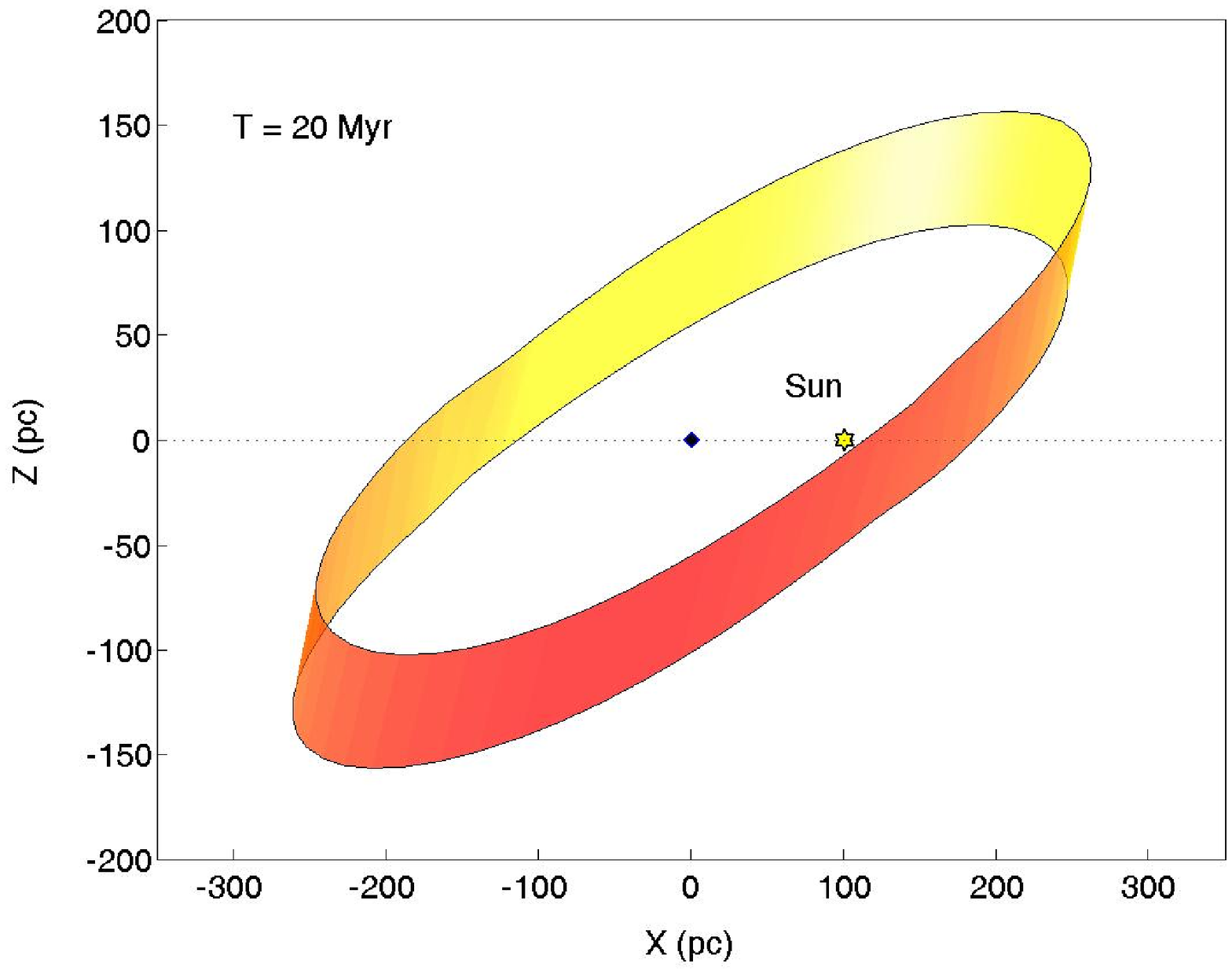}}
\centerline{\includegraphics[width=0.45\linewidth]{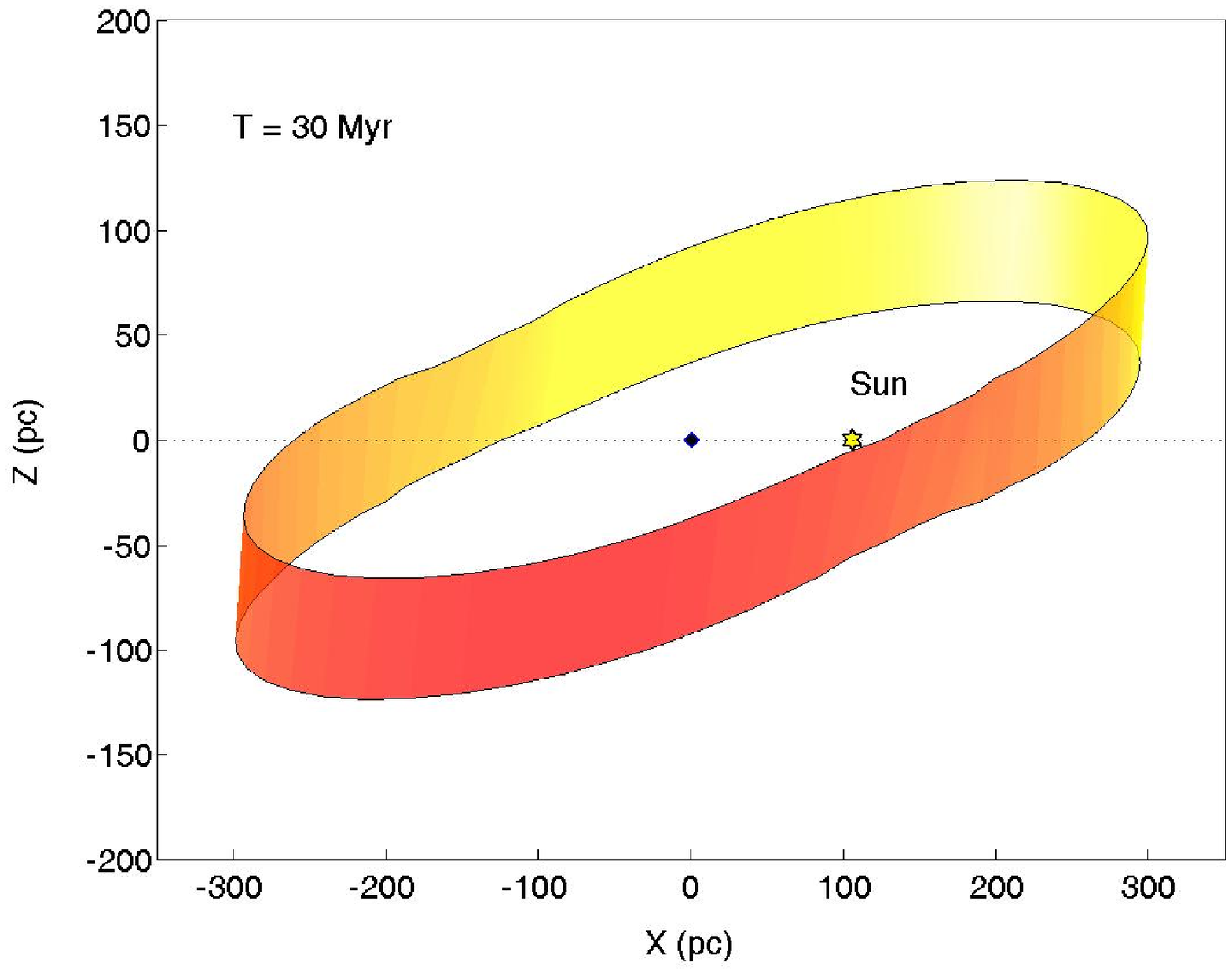}
\includegraphics[width=0.45\linewidth]{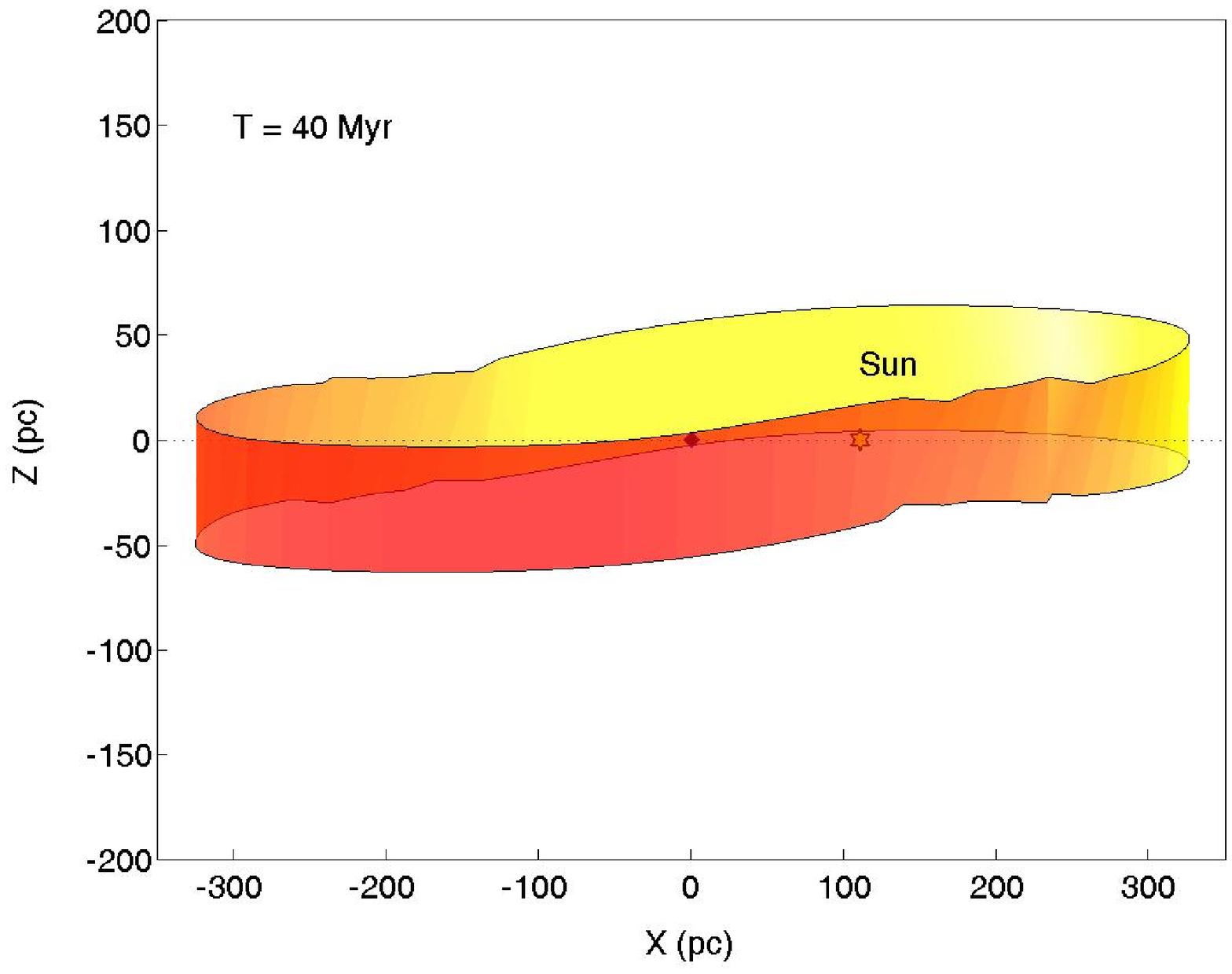}}
\caption{The Gould Belt evolution as seen at different epochs
after the outburst, in a plane perpendicular to the Galactic
plane, centred on the Belt centre. The x axis points to the
Galactic centre and the location of the Sun is marked by an
asterisk. The Belt expands and is further distorted by the
combined effects of the Galactic differential rotation,
interstellar density gradients above the Galactic plane, and the
gravitational pull of the Galactic disc. The plot for 30 Myr is
representative of the present-day geometry.}\label{fig:Belt:evol}
\end{figure*}

Figure \ref{fig:Belt:evol} shows the maximum-likelihood evolution
of the Belt rim, as seen at different epochs in a plane
perpendicular to the Galactic one. The current Belt geometry is
close to that depicted in the T = 30 Myr plot and the last plot
illustrates how the Belt may fall back onto the Galactic disc
within 10 to 15 Myr. The height that best fits the data is $H = 60
\pm 1$ pc. The expansion proceeds faster at higher $z$ altitudes
because of the reduced ambient interstellar density. The inclined
Belt starts by rapidly expanding out of the Galactic disc. Its
shape gradually narrows into an ellipse that precesses because of
the swept-up momentum from the interstellar gas and its
differential rotation. The inclination also decreases with time
and the disc gets clearly warped because of the gravitational pull
from the Galactic disc.

To zeroth order, the expansion radius R(t) is well described by
momentum conservation in a uniform medium of density $\rho$:
$\frac{d}{dt}(\pi \rho H R^{2} \dot{R}) = 0$. The resulting
power-law dependance $R \propto t^{1/3}$ is found in very good
agreement with the evolution of the average size, R(t)=(a+b)/2,
obtained from the model after the first 200 kyr.

The best fit yields a position and velocity field for the present
Belt that are in good agreement with the cloud data, as
illustrated in Figures \ref{fig:lonlat}, \ref{fig:lonvit}, and
\ref{fig:lonlatvit}. Figure \ref{fig:lonlat} shows the trace of
the Belt across the sky in longitude and latitude. Apart from a
few points, which have not been removed from the sample to avoid
biases, the modelled Belt geometry nicely fits the data. Its
angular width increases at low longitudes where the rim is closer
and the varying angular width matches the angular dispersion seen
in the nearby clouds. Figure \ref{fig:lonvit} shows the cloud
distribution in longitude and velocity, as well as the velocity
field of the present rim. The slow expansion of the Belt to date
reproduces the double-peaked velocity pattern over all longitudes.
The velocity pattern clearly differs from the $v_{LSR} =
A.d.sin(2l).cos^{2}(b)$ dependence expected from a pure
differential rotation. The difference reflects the tilt of the
Belt and its expansion. The peak-to-peak velocity amplitude is,
however, not quite satisfactory, even in the face of the large
intrinsic dispersion in the cloud data. This aspect will be
addressed in section \ref{ssec:velocity}. Figure
\ref{fig:lonlatvit} shows that the model reasonably reproduces the
complex data distribution in the longitude, latitude, and velocity
phase space.
\begin{table}
\caption{Initial and present characteristics of the Gould Belt}
\label{tab:fitparam}
\begin{tabularx}{0.7\linewidth}{rrr}
\hline
                    & initially                & today          \\
\hline
$l_{\Omega}$ ($\degr$)  & $290.1 \pm 1.1$          & $296.1 \pm 2.0$\\
$\varphi$ ($\degr$)     & $35.2 \pm 0.3$           & $17.2 \pm 0.5$ \\
$l_{c}$ ($\degr$)       & $129.7 \pm 2.0$          & $180.4 \pm 2.0$\\
$d_{c}$ (pc)        & $145 \pm 4$              & $104\pm 4$     \\
$a$ (pc)            &                          & $373 \pm 5$    \\
$b$ (pc)            &                          & $233 \pm 5$    \\
\hline \\
$\tau$ (Myr)        & $E_{i}$ ($10^{45}$ J)    & $H$ (pc)       \\
$26.4 \pm 0.4$      & $1.0 \pm 0.1$            & $60 \pm 1$     \\
\hline
\end{tabularx}
\end{table}
\subsection{Geometry}
\begin{figure*}
\includegraphics[width=\linewidth]{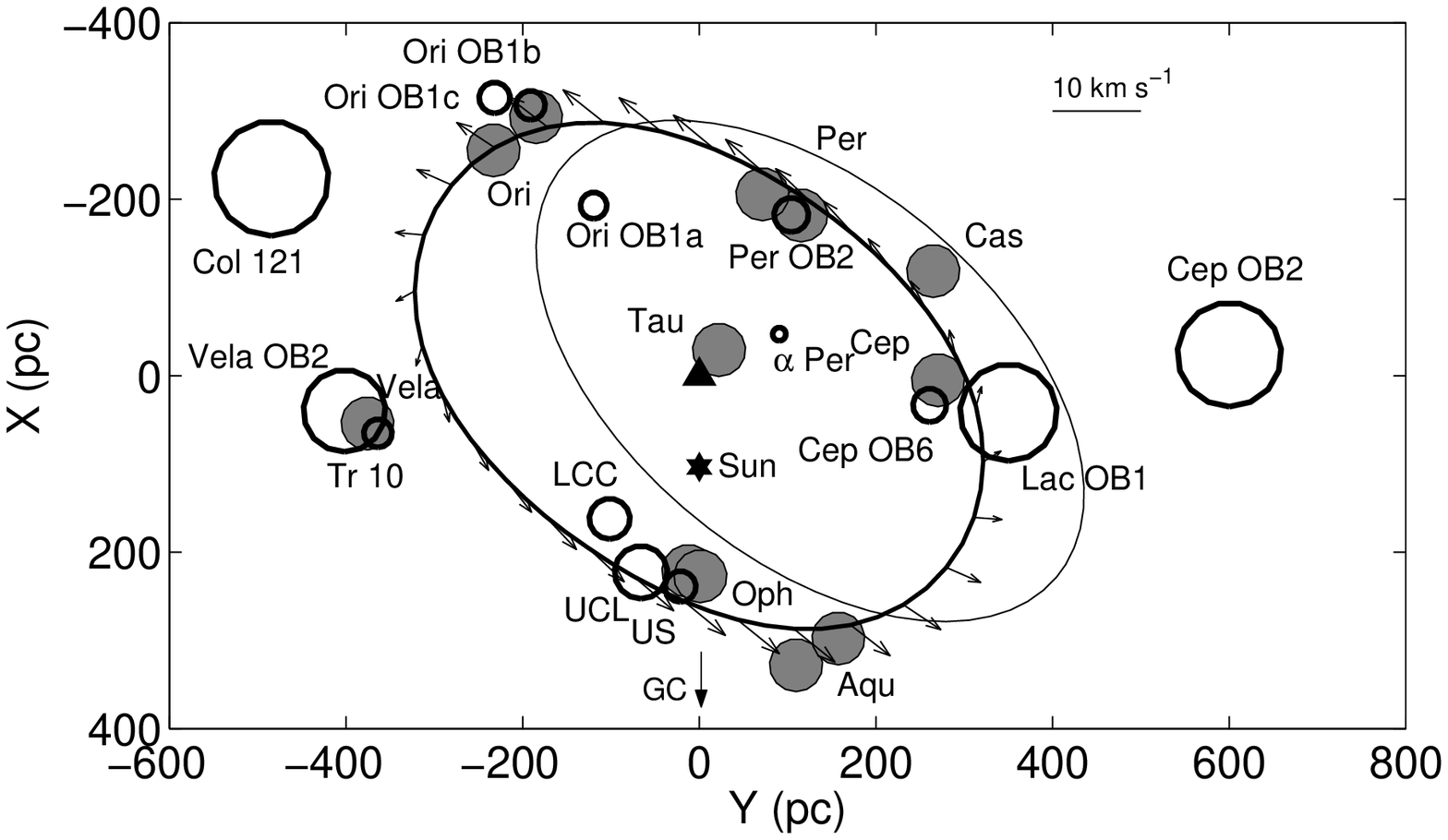}
\caption{Present position of the Gould Belt projected on the
Galactic plane. The x and y axes point to the Galactic Centre and
in the direction of the Galactic rotation, respectively. Both are
centred on the Belt centre. The velocity field outlines the Belt
expansion with respect to the Local Standard of Rest. Nearby OB
associations are plotted as thick circles using \textit{Hipparcos}
estimates of their distance and dimensions from
\citet{de_zeeuw_1999_art}. The shaded circles mark the location of
the main nearby H$_{2}$ cloud complexes. The thick and thin
ellipses note the Belt rim as obtained in this work and earlier
from the HI data by \citet{olano_1982_art}. The triangle and star
note the Belt centre and the Sun, respectively.}
\label{fig:Belt:GP:proj}
\end{figure*}
The present ellipse has a semi-major axis $a = 354 \pm 5
\;\mathrm{pc}$ and a semi-minor axis $b = 232 \pm 5
\;\mathrm{pc}$, in good agreement with the dimensions of 360 by
210 pc and 341 by 267 pc derived by \citet{olano_1982_art} and
\citet{moreno_1999_art} from the sole HI data when modelling the
expansion of a 2D ring in the Galactic plane or a 3D superbubble.
The maximum-likelihood fit also yields a present inclination
$\varphi = 17.2\degr \pm 0.3\degr$ on the Galactic plane that
nicely compares with the stellar estimates. Inclinations of
22.3$\degr$, $16\degr-22\degr$, and $20\degr$ have been found from
the massive-star population by \citet{comeron_1994a_art}, \citet{
torra_2000a_art}, and \citet{olano_2001_art}, respectively. A
lower value of $12.5\degr$ was derived from the molecular clouds
\citep{taylor_1987_art}. The difference with the present result
can be attributed to a different cloud dataset. Taylor et al. did
not select high-latitude dark clouds that are shown below to be
consistent with the Belt shell. An inclination near $20\degr$
supports a strong connection between the Gould Belt and the Local
Bubble (or Chimney) since the axis of the latter appears to be
perpendicular to the Belt plane \citep{sfeir_1999_art}.
\begin{figure*}
\includegraphics[width=0.9\linewidth]{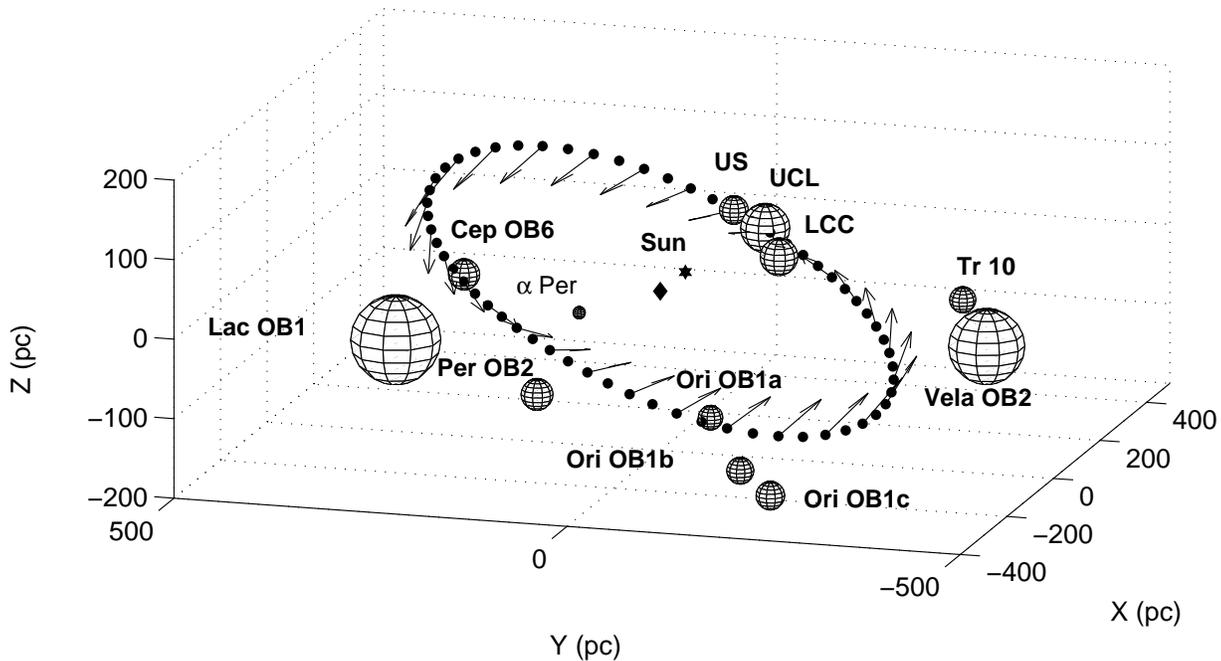}
\caption{3D view of the present Belt wave and its velocity field
(with respect to the LSR) amid the local OB associations that are
marked as spheres with a radius equal to their size
\citep{de_zeeuw_1999_art}. The diamond and star note the Belt
centre and the Sun, respectively.}\label{fig:Belt:3D}
\end{figure*}

We find a position and orientation that significantly differ from
previous estimates. The Belt centre is found at a distance
$d_{centre}^{today} = 104 \pm 4$ pc from the Sun, in the direction
$l_{centre}^{today} = 180\degr \pm 2\degr$. More distant centres,
well into the second quadrant, were proposed from the HI data by
\citet{olano_1982_art} and \citet{moreno_1999_art} (131 pc and 173
pc away, towards $166\degr$ and $117\degr$, resp.). The ellipse
axes also point to a different direction. Figure
\ref{fig:Belt:GP:proj} gives the current Belt position projected
onto the Galactic plane. The change in orientation can be
attributed to the refined distance information used here and to
the presence of new major nearby H$_{2}$ complexes, such as
Cepheus, Cassiopeia, and Polaris. \citet{taylor_1987_art} did not
select their dark clouds as Belt members on a direction basis, but
their combined direction, velocity, and distance appear here to be
quite consistent with their being part of the expanding shell in
the second quadrant (see Figures \ref{fig:lonlat} and
\ref{fig:lonvit}).

Finally, the best fit yields an ascending node longitude
$l_{\Omega} = 296.1\degr \pm 2.0\degr$ that is $\sim 10\degr$
higher than the values obtained from the stars ($l_{\Omega} =
284.5\degr$ from the massive ones \citep{comeron_1994a_art} and
$l_{\Omega} = 282\degr$ from the young low-mass stars
\citep{guillout_1998b_art}). The difference possibly reflects the
time-lag between the stellar birth epoch and the slowly precessing
rim today.

As shown in Figure \ref{fig:Belt:GP:proj}, the position of the
shell nicely coincides with most of the nearby OB associations
such as Per OB2, Ori OB1a \& OB1c, LCC, UCL, US, Cep OB6, Tr 10,
Vela OB2, the positions of which are known from \textit{Hipparcos}
measurements \citep{de_zeeuw_1999_art}. Col 121 and Cep OB2
clearly lie outside the Belt. Projections can be misleading. The
3D view displayed in Figure \ref{fig:Belt:3D} illustrates that Lac
OB1, often quoted in relation to the Belt, lies on the wrong side
of the Galactic plane to be part of it, even when taking into
account the Belt thickness of 60 pc. The old age of the
association, of the order of 12--16 Myr \citep{de_zeeuw_1999_art},
also confirms that Lac OB1 is not related to the Belt.
\begin{table*}
\caption{Maximum likelihood values reached for different scenarii.
P is the probability that random fluctuations from a given model
fit the data as well as the expansion model. Characteristic Belt
parameters are given for a selection of scenarii.}
\begin{tabularx}{1.0\linewidth}{lcccccccc}
\hline
               & $v_{rot}$ (km/s)  & $\ln(L_{max})$  & $P$ & $\tau (Myr)$ & $E_{i}$ ($10^{45}$ J) & $\varphi_{today} (\degr)$ & $a_{today}$ (pc) & $b_{today}$ (pc) \\ \hline
expansion      & 0                 & -625.5 &        & $26.4 \pm 0.4$ & $1.0 \pm 0.1$ & $17.2 \pm 0.5$ & $373 \pm 5$ & $233 \pm 5$ \\
rotation       & 5                 & -626.2 & 25.4\% \\
rotation       & 50                & -625.9 & 35.9\% \\
rotation       & -50               & -625.6 & 77.7\% & $26.4 \pm 0.5$ & $1.1 \pm 0.1$ & $17.3 \pm 0.5$ & $372 \pm 6$ & $232 \pm 6$ \\
rotation       & 100               & -626.0 & 34.8\% \\
rotation       & -100              & -625.5 & 100\%  & $26.4 \pm 0.5$ & $1.0 \pm 0.1$ & $17.1 \pm 0.5$ & $371 \pm 5$ & $232 \pm 5$ \\
plane crossing & 0                 & -936.3 & $\ll$  & $51.8 \pm 1.0$ & $0.6 \pm 0.06$ & $15.1 \pm 0.7$ & $304 \pm 6$ & $219 \pm 6$ \\
fragmentation  & 0                 & -667.6 & 5e-20  & $28.9 \pm 0.6$ & $0.3 \pm 0.03$ & $5.7 \pm 0.5$ & $142 \pm 5$ & $72 \pm 5$ \\
\hline
\end{tabularx}
\label{tab:likelihood}
\end{table*}
\subsection{Effects of an initial rotation, shell fragmentation and Galactic plane crossing}
\label{ssec:velocity}
The velocity dispersion of the data points in Figure
\ref{fig:lonvit} is large, but consistent with the cloud to cloud
dispersion observed in the interstellar medium. A larger
peak-to-peak amplitude in the shell velocity would, however,
better represent the data. The influence of the initial
conditions, the Belt thickness, the distance information, and the
interstellar density profiles have been investigated to reduce
this discrepancy, but none of these options leads to a significant
increase in velocity amplitude within sensible limits. The only
sensitive parameter is the Oort's constant A, but unrealistic
values beyond 17 km s$^{-1}$ kpc$^{-1}$ are required to slightly
improve the fit. Other scenarii have been studied.

One possibility was to allow the Belt to cross the Galactic plane
before reaching its present orientation. The effect of an initial
rotation has also been considered by adding a tangential component
to the initial expansion velocity. A more realistic braking of the
shell has been introduced in the late stages of the evolution by
gradually applying a drag force on the shell elements when their
velocity falls into the 30 to 10 km s$^{-1}$ interval. In
parallel, shell fragmentation has been implemented by increasing
the porosity to the interstellar flow. The porosity coefficient
$\alpha_{p}$ varies linearly from a pure snowplough case above 20
km s$^{-1}$ ($\alpha_{p} = 0$) to a pure drag without accretion
below 5 km s$^{-1}$ ($\alpha_{p} = 1$). The drag force is given by
$\overrightarrow{dF} =
-\alpha_{p}\;C_{x}\;\rho_{gas}\;(\overrightarrow{v}'\cdot\overrightarrow{dS})
\;\overrightarrow{v}'$ where $C_{x}$ denotes the friction
coefficient. A value of 0.5 was adopted corresponding to turbulent
friction on a sphere. $\overrightarrow{dS}$ is an elementary
surface of the shell front, pointing outwards, and
$\overrightarrow{v}'$ gives the relative velocity of
$\overrightarrow{dS}$ with respect to the interstellar medium.

The maximum-likelihood values reached for these different scenarii
are given in Table \ref{tab:likelihood}. Probabilities for random
fluctuations from a model to fit the data as well as the original
expansion model are estimated from the likelihood ratio
$-2\;\ln\lambda = -2\;\ln(\mathcal{L}_{max\,
model}/\mathcal{L}_{max\, exp})$ between the two models. This
ratio follows a $\chi^{2}_{1}$ distribution
\citep{eadie_1971_book}. The corresponding probabilities are given
in Table \ref{tab:likelihood}. It can be seen that a scenario
including a late shell fragmentation or a Galactic plane crossing
can hardly reproduce the data. The improvement in the quality of
the fit of the expansion model over these scenarii is very
significant.

Because of the reduced deceleration of the fragmented shell, a
lower initial velocity is needed to match the low cloud velocities
and the small size and low inclination resulting for the evolved
Belt in this case explain the much poorer fit.

\begin{figure*}
\centerline{\includegraphics[width=0.45\linewidth]{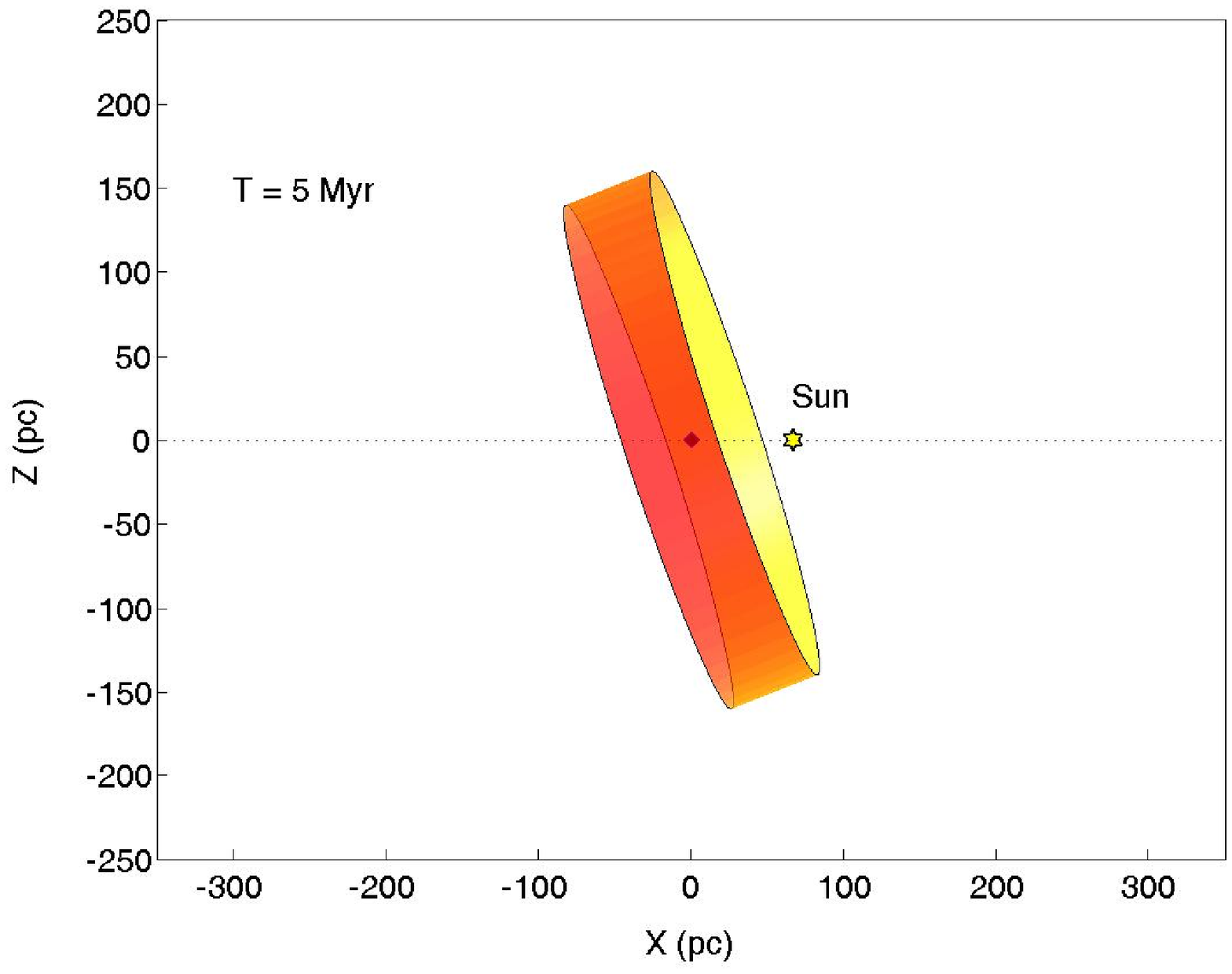}
\includegraphics[width=0.45\linewidth]{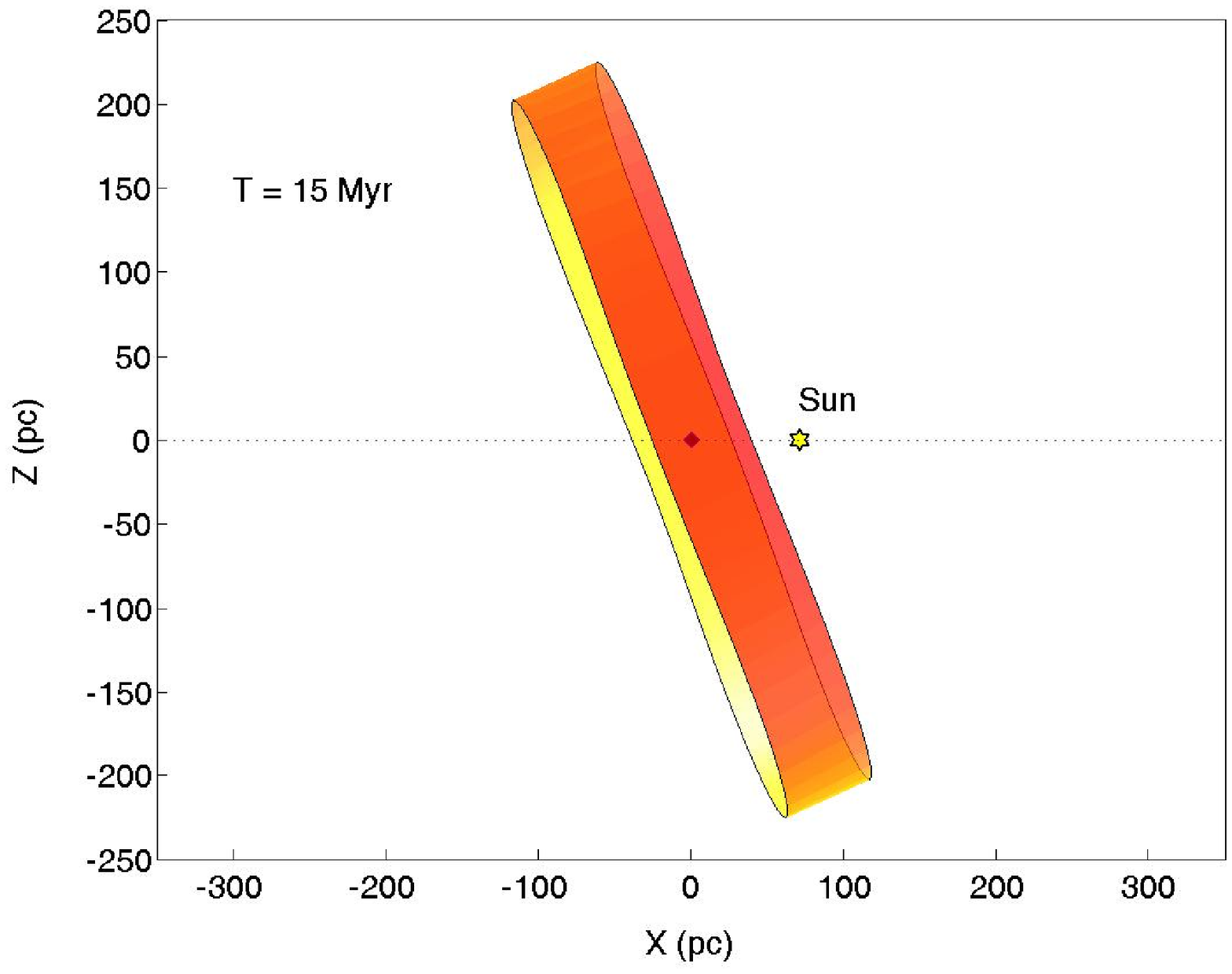}}
\centerline{\includegraphics[width=0.45\linewidth]{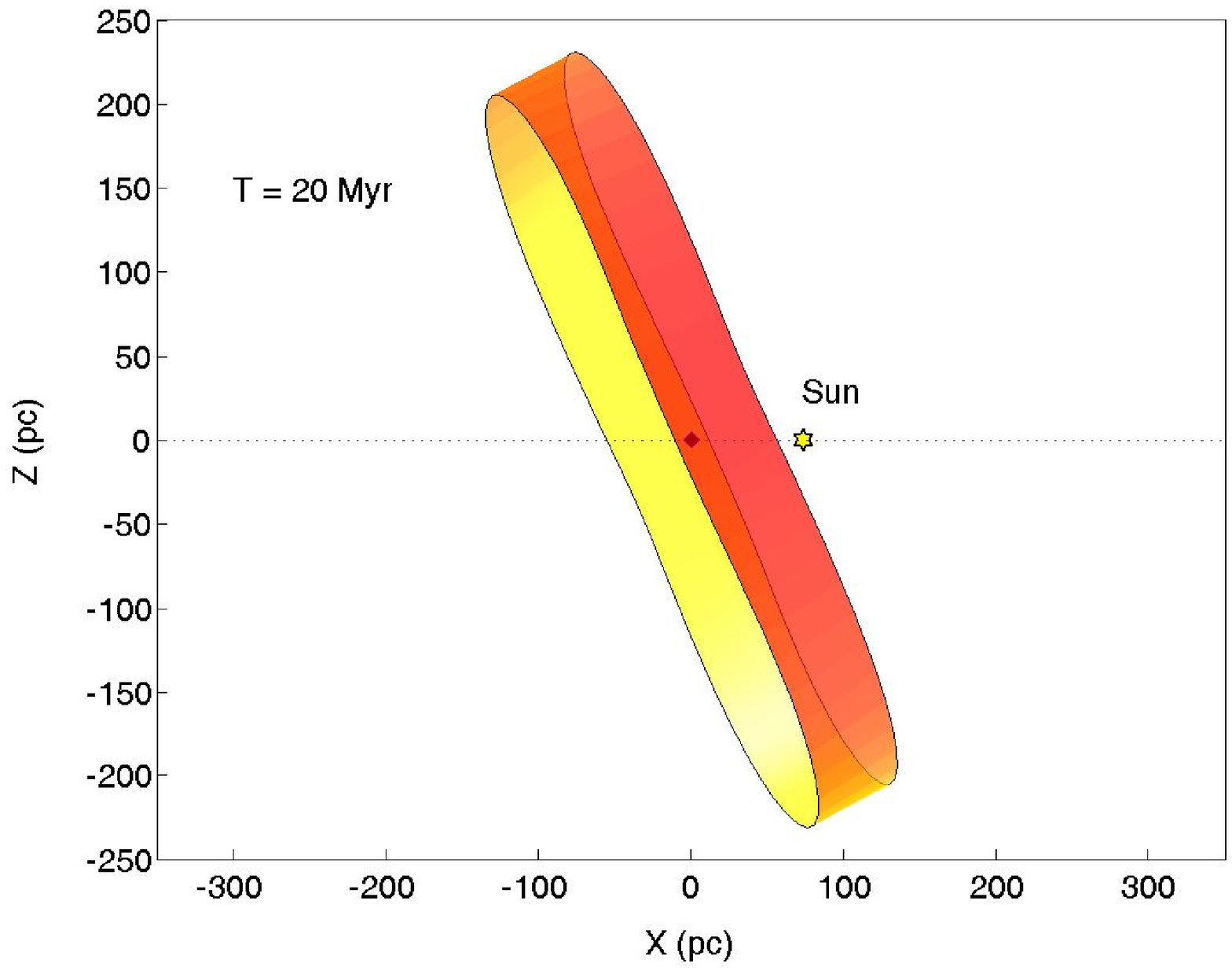}
\includegraphics[width=0.45\linewidth]{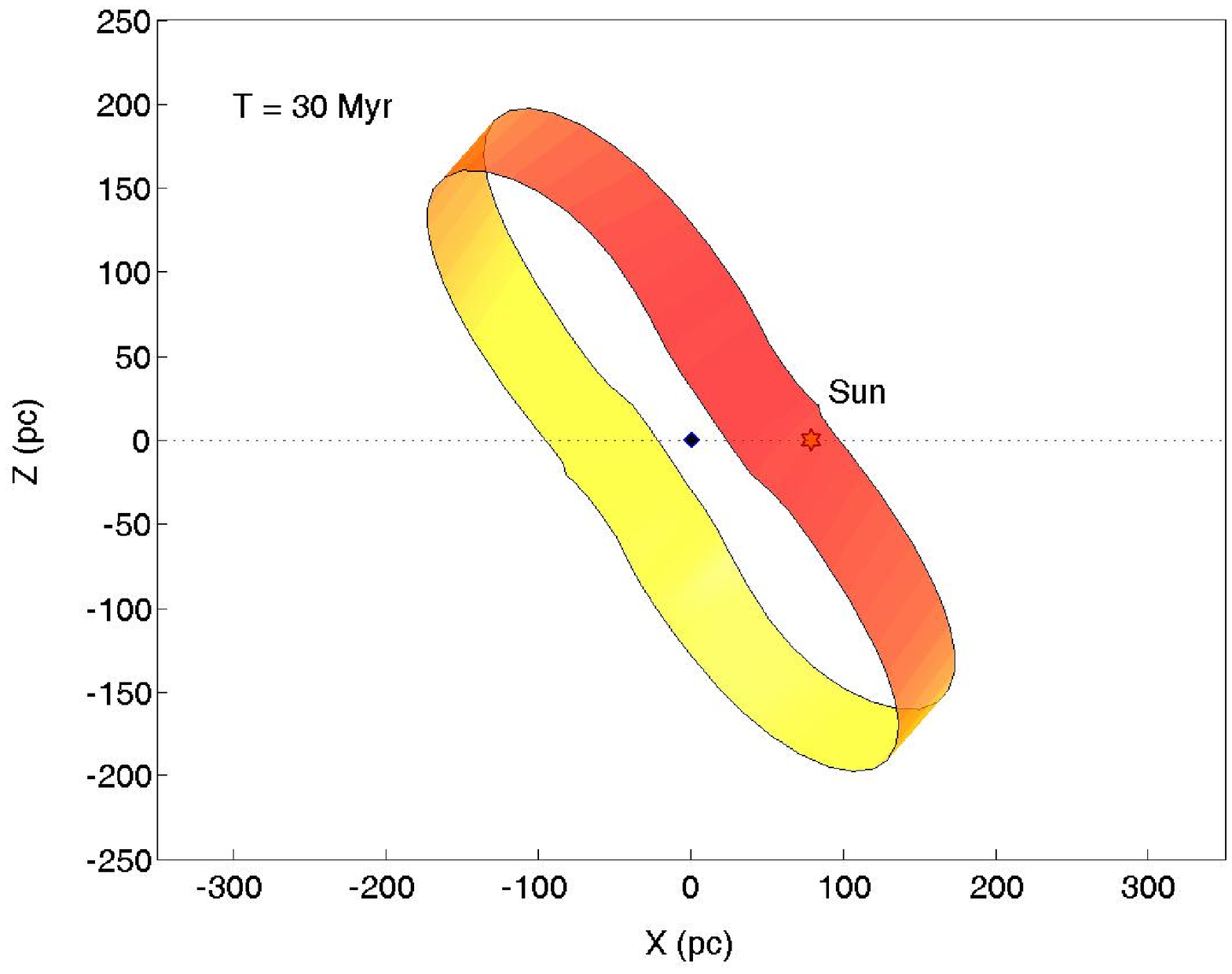}}
\centerline{\includegraphics[width=0.45\linewidth]{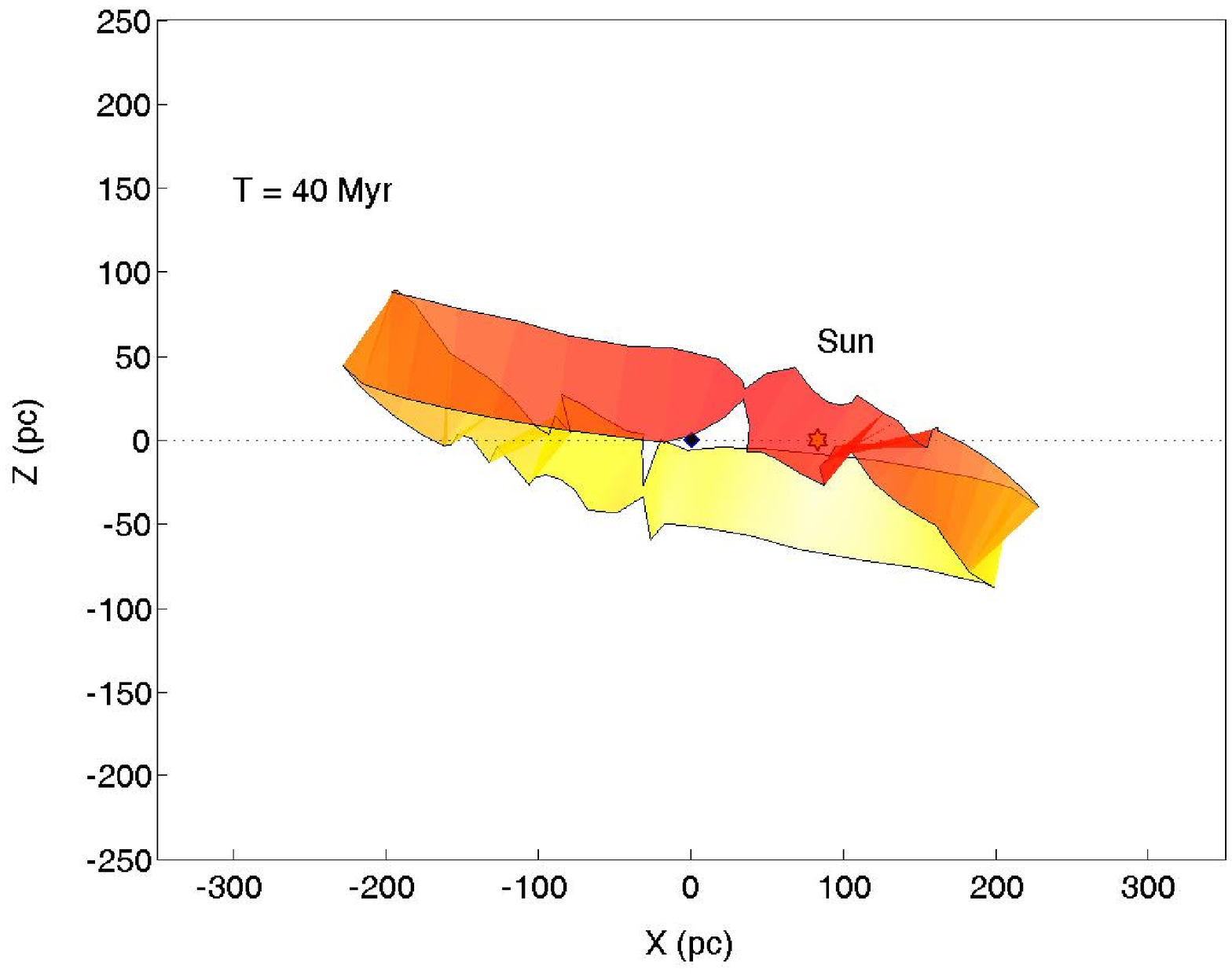}
\includegraphics[width=0.45\linewidth]{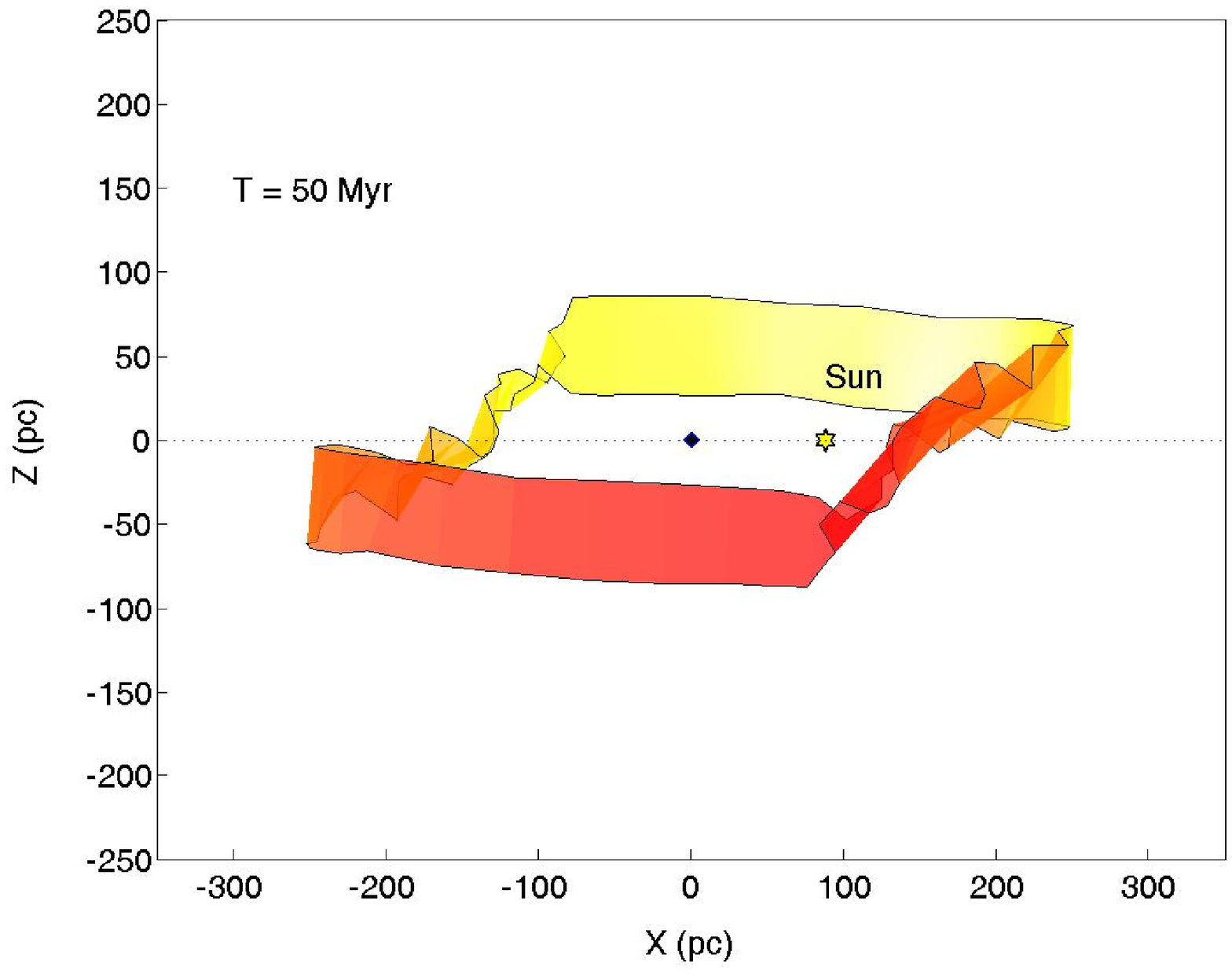}}
\caption{The Gould Belt evolution as seen at different epochs
after the outburst (see Figure \ref{fig:Belt:evol:cross}),
assuming a crossing of the Galactic plane. The plot for 50 Myr is
representative of the present-day geometry.}
\label{fig:Belt:evol:cross}
\end{figure*}

The crossing of the Galactic plane results in an hour-glass like
distorsion of the shell because of the interstellar density
gradient implying different expansion velocities in and out of the
Galactic disc. A maximum-likelihood solution could, however, be
found. It requires a larger initial inclination to boost the Belt
to larger altitudes and make use of the increased gravitational
pull towards the plane to cross it and reach a 20$\degr$ tilt on
the other side. A lower initial kinetic energy is needed because
of the extra energy provided by the disc torque. The evolved Belt
disc ends up being quite warped. As expected from Figure
\ref{fig:Belt:evol:cross}, the resulting fit is much poorer than
without plane crossing. This is confirmed by a dramatic decrease
in the maximum likelihood. This scenario requires a substantially
longer time span of $51.8 \pm 1.0$ Myr.

The fit seems rather insensitive to an initial rotation, even for
large rotation velocities comparable to the expansion one. The age
and energy estimates, as well as the final geometry, are not
affected by an initial rotation. This set of results show that the
dominant factor in the shell dynamics and final velocity
distribution is the momentum accreted from the interstellar gas.

\subsection{Age of the Belt}
The best value found for the age of the Belt, $\tau = 26.4 \pm
0.4$ Myr, is comparable to previous estimates based on the gas
dynamics (31-36 Myr in the 2D model of \citet{olano_1982_art}, 23
Myr and 15.5 Myr in the 3D superbubble model of
\citet{moreno_1999_art}, without and with interstellar pressure).
The shorter timescales involved in the 3D models may be due to the
faster expansion in the rarefied medium away from the Galactic
disc in the early stages. So, the age estimate is sensitive to the
interstellar stratification. The dynamics of the stellar system,
mainly its vertical oscillations and estimates of the Oort
constants, imply an age of $34 \pm 3$ Myr
\citep{comeron_1999_art}. Taken together, these various
measurements suggest a dynamical age of the Belt of order 30 Myr.

Yet, stellar ages point to a twice longer timescale for the Belt
system. Determining the age of the massive stars from photometric
measurements, \citet{westin_1985_art} proposed an upper limit of
60 Myr, and \citet{torra_2000a_art} showed that 60 to 66\% of the
stars younger than 60 Myr and closer than 600 pc belong to the
Belt. The X-ray luminosities of $10^{30.5 \pm 0.5}$ erg/s of the
young low-mass stars correspond to ages between 30 and 80 Myr
\citep{guillout_1998b_art}. One difficulty in determining the Belt
age from the stars is to reliably separate the Belt and the
Galactic populations. Another comes from deriving the photometric
age without properly taking into account the stellar rotation.
\citet{figueras_1998_art} indeed found that photometric ages of
rotating B7-A4 stars are overestimated by 30 to 50\% on average.
This important bias would also affect the highly rotating massive
stars and would help reduce the discrepancy between the dynamical
age of $\sim 30$ Myr and the stellar age of $\sim 60$ Myr. On the
other hand, an origin in a single explosive event or continuous
energy injection would imply different dynamical ages. As
discussed in the next section, gradual injection would further
reduce the dynamical timescale. One way to reduce the dynamical
and stellar age discrepancy is to allow the Belt to cross the
Galactic plane before reaching its present orientation. The much
poorer fit for this scenario, however, does not support this
possibility.
\subsection{Input energy}
The expansion results from an initial kinetic energy input of
$(1.0 \pm 0.1)$ $10^{45}$ J which is comparable, but slightly
higher than the energy of 6.1 10$^{44}$ J required in the 2D
expansion model of \citet{olano_1982_art}. This is due to the
extra amount of work needed to expand against the gravitational
pull of the Galactic disc in the early phases. An equivalent
energy of 6 10$^{44}$ J is required for the 3D expansion of a
superbubble, in a 3 times lower interstellar density, but against
ambient pressure \citep{moreno_1999_art}. The 10 times lower value
they found for pressureless expansion is due to their choice of a
very low density. An initial energy of $10^{45}$ J is typical of
the energy deposit from multiple supernovae inside a young stellar
cluster, but it would require a series of explosions within a
rather short period. The energy budget is also typical of a
hypernova powering a $\gamma$-ray burst event. Exploring whether
the asymetrical explosion can explain the Belt tilt is under
study. The energy budget can also be accounted for by the
potential energy of high-velocity clouds falling on the Galactic
disc \citep{comeron_1992_art, comeron_1994a_art}.

Continuous energy injection by multiple supernovae and expansion
into a highly inhomogeneous medium would yield a larger dynamical
range in velocity. Injecting energy gradually would, however,
accelerate the Belt expansion. To zeroth order approximation,
expansion in a uniform medium of density $\rho$, with constant
power injection $P_{inj}$, can be modelled by $\frac{d}{dt}(\pi
\rho H R^{2} \dot{R}) = 2\pi H R p_{int}$, where $p_{int}$
represents an internal source of pressure to account for the
energy deposit. Its mechanical power is $2\pi H R p_{int} \dot{R}=
P_{inj}$. Solving for power-law expansion ($p_{int}
\propto{R^{-\gamma}}$ and $R \propto{t^{n}}$), one finds a faster
expansion $R \propto{t^{3/4}}$ for constant power injection than
from a single explosion ($R \propto{t^{1/3}}$). The faster
evolution would further reduce the dynamical age with respect to
the stellar age, and also reduce the Belt eccentricity with
respect to the location of the clouds and OB associations. The
complex interplay between the power injection and momentum
accretion is being investigated.
\subsection{Mass accretion}
\begin{figure}
\includegraphics[width=0.9\linewidth]{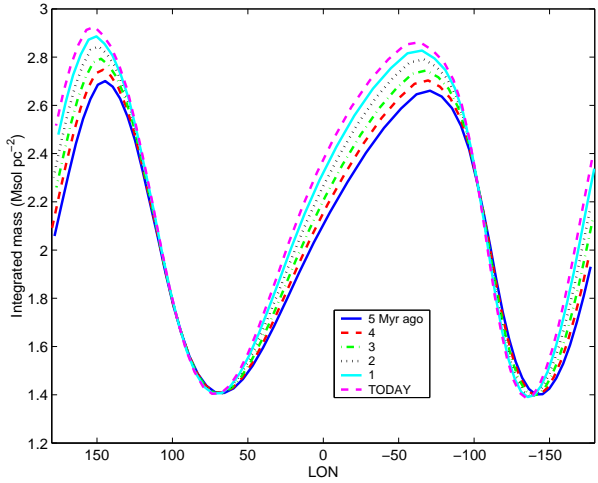}
\caption{Total integrated mass along the Gould Belt rim, in solar
masses per pc$^{2}$, in the snowplough case, at five different
epochs over the past 5 Myr.} \label{fig:mass:distrib}
\end{figure}
The swept-up mass in the evolved shell is found to be 2.4 10$^{5}$
M$_{\odot}$. The larger value of 1.2 10$^{6}$ M$_{\odot}$ proposed
by \citet{olano_1982_art} results from his choice of a 20 \%
larger density in the Galactic plane and the absence of vertical
gradients and expansion away from the plane. The estimate of 3.3
10$^{5}$ M$_{\odot}$ found by \citet{moreno_1999_art} in the 3D
supershell is close to our result. The 3 times lower density value
they adopted is compensated by the lack of vertical gradients.

Figure \ref{fig:mass:distrib} shows the integrated mass
distribution along the Belt over the past 5 Myr. The mass
accumulated in the snowplough case is lowest in the directions
$35\degr < l < 100\degr$ and $-155\degr <l < -110\degr$. These
regions in the first and third quadrants are indeed relatively
free of major cloud complexes (\textit{c.f.} Figure
\ref{fig:Belt:GP:proj}). Conversely, the swept-up mass is highest
in the $-100\degr < l < 22\degr$ and $-165\degr < l < 115\degr$
intervals that encompass most of the main nearby complexes. The
nodal line of maximum accretion is close to the Belt nodal line,
illustrating the importance of the Belt tilt and of the vertical
density gradient. In other words, the Belt accretes more near the
Galactic disc. Figure \ref{fig:mass:distrib} illustrates the Belt
precession, of the order of a few degrees per Myr, that can be
attributed to the Galactic differential rotation.
\subsection{The Belt wave and the OB associations}
\begin{figure}
\includegraphics[width=0.9\linewidth]{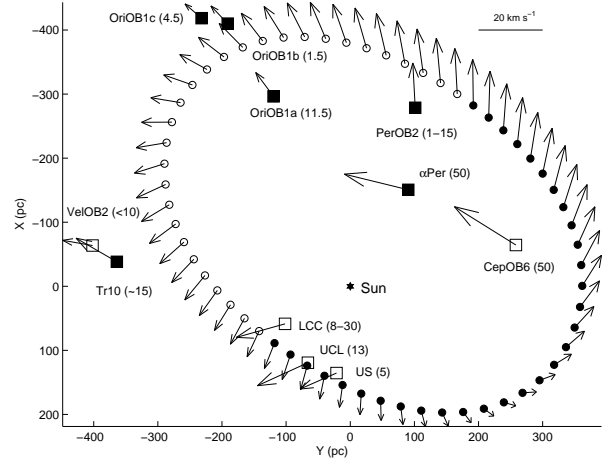}
\caption{Present-day velocities (projected on the Galactic plane)
of the nearby OB associations (squares) \citep{de_zeeuw_1999_art},
and of stars that would have been born in the expanding Belt shell
10 Myr ago. Their position (circles) and velocity today have been
computed after ballistic motion in the Galactic potential. All
velocities are corrected for solar motion \textit{and} Galactic
rotation. Filled and open symbols mark negative and positive
vertical $v_{z}$ velocities, respectively. The presumed age for
each OB association is given in Myr. The Sun is marked by a star.}
\label{fig:OB10Myr}
\end{figure}
The velocity field of the nearby young stellar groups (see Figure
5 in \citet{lindblad_1997_proc}) suggests a stream-like motion
spreading out of the Belt region. Because of the frictionless
motion of the stars in the Galactic potential, one expects older
stellar groups, born in a faster expanding Belt shell, to travel
farther out than younger groups born from a slowed-down Belt. The
opposite situation is observed in Figure \ref{fig:OB10Myr} where
younger OB associations are found at larger radii than older ones
(notice in particular the LCC/UCL/US and the Ori1 a/b/c
sequences). The present-day positions and velocities of stars born
10 Myr ago from the expanding shell have been computed from their
ballistic motion in the Galactic potential. Note that velocities
were corrected for solar motion (using $U_{x\odot} = 10$ km
$s^{-1}$, $V_{y\odot} = 5$ km $s^{-1}$, and $W_{z\odot} = -7$ km
$s^{-1}$) \textit{and} for Galactic rotation (using Oort's
constants $A = 14.8$ km s$^{-1}$ kpc$^{-1}$ and $B = -12.4$ km
s$^{-1}$ kpc$^{-1}$) in order to allow a direct comparison with
Figure 29 from \citet{de_zeeuw_1999_art}. It implies that
velocities are considered with respect to the Local Standard of
Rest at the position of the object considered. One expects the
computed present-day positions to be systematically too far out
because the Belt shell model has been fitted against the current
position of the OB associations, not against their birth positions
several Myr ago. There is, however, no convincing correlation
between the observed and expected velocity fields, particularly in
the vertical velocity component due to the gravitational pull from
the Galactic disc. The current data suggests that the Belt wave
triggers star formation when overtaking a cloud and that the
association average velocity does not relate to that of the
progenitor cloud.
\section{Conclusion}
\label{sec:conclusion}
The Gould Belt expansion has been modelled in 3D and has been
fitted against the current direction and velocity of the H$_{2}$
and HI clouds that are found in the solar neighbourhood. Distance
measurements have also been used for a subset of clouds. Due to
the combined effects of the Galactic differential rotation, its
gravitational pull, and the interstellar density gradients, the
present Belt section has evolved into a slightly warped ellipse,
with semi-axes of $(373 \pm 5)$ pc and $(233 \pm 5)$ pc, and an
inclination of $17.2\degr \pm 0.5\degr$. Its centre is currently
located $(104 \pm 4)$ pc away from the Sun, in the direction
$l_{centre} = 180.4\degr \pm 2.0\degr$. The thickness that best
fits the data is $H = (60 \pm 1)$ pc. While these characteristics
nicely compare with previous estimates, a different Belt
orientation has been found that is driven by the presence of new
major H$_{2}$ complexes and the revised distance information used
here from Hipparchos measurements. The Belt position and
orientation are found to coincide with the main OB associations
and molecular clouds in our vicinity.  A larger velocity amplitude
would better represent the gas pattern, but it turns out to be
insensitive to all model ingredients but for the Oort constant A.
Other scenarii have been tested including shell fragmentation in
the late stages, an initial rotation, or the crossing of the
Galactic disc. They did not yield an increase in the final
dynamical range in velocity. Much poorer fits to the data were
obtained for the fragmented shell and the disc crossing cases. The
Belt evolution has been compared to the kinematics of the nearby
OB associations. Its expansion appears to have no influence on the
observed average stellar motions. On the other hand, the swept-up
mass distribution along the Belt rim, for a total of 2.4 $10^{5}$
M$_{\odot}$, is reasonably consistent with the cloud distribution.
The younger OB associations are surprisingly found farther away
from the Belt centre than the older ones.

The initial kinetic energy of $E_{i} = (1.0 \pm 0.1)$ 10$^{45}$ J
amounts to that of 10 supernovae, as previously proposed by
several authors. The dynamical age of $(26 \pm 0.4)$ Myr is
equivalent to earlier estimates based on the gas expansion, but it
is only half of the 60 Myr age derived from photometric stellar
ages. This discrepancy does not result from the choice of a single
explosion at the origin. Continuous energy injection would
increase the discrepancy by further reducing the dynamical age.
The age estimate is not sensitive to an initial rotation either.
Allowing one crossing of the Galactic disc before reaching the
present inclination implies a dynamical age of 52 Myr in better
agreement with the stellar ages, but the very poor fit does not
support this possibility. Important biases in the photometric
derivation of stellar ages for rapidly rotating stars have been
reported that could help solve this discrepancy.
\begin{acknowledgements}
The authors would like to thank J. de Bruijne and A. Brown for
kindly providing data on the OB associations. We gratefully
acknowledge the referee, A. Blaauw, for his helpful comments and
discussions. We also thank J.-P. Chi\`{e}ze for useful discussions
on hydrodynamical issues.
\end{acknowledgements}

\bibliographystyle{aa}
\bibliography{PerrotGrenier02rev}

\newpage
\appendix
\section{CLUMPFIND results}
For each clump detected with \textsc{clumpfind}, the position and
velocity of the corresponding centroid are given in the following
Tables. For known molecular complexes, distances that were used in
the likelihood function (\textit{c.f.} section
\ref{ssec:likelihood}) are also indicated.
\begin{table}
\caption{HI clumps}
\begin{tabularx}{0.8\linewidth}[h]{rrrrl} \hline
 l      & b         & v$_{LSR}$     & D             & complex                 \\
 (deg)  & (deg)     & (km/s)        & (pc)          &                         \\ \hline
98,5    &   8,5     &   -1,030529   &   ?           &                         \\
106     &   7       &   -6,18318    &   ?           &                         \\
16,5    &   13,5    &   1,030531    &   ?           &                         \\
17,5    &   12,5    &   1,030531    &   ?           &                         \\
102,5   &   10      &   -7,21371    &   ?           &                         \\
-169,5  &   -13,5   &   4,122121    &   340         & Ori                     \\
-159    &   -9,5    &   7,213711    &   500         & Ori$^{\mathrm{a}}$      \\
-173    &   -14,5   &   3,091591    &   340         & Ori                     \\
-157,5  &   -13,5   &   6,183181    &   500         & Ori$^{\mathrm{a}}$      \\
-140    &   -12,5   &   4,122121    &   ?           &                         \\
-168,5  &   -13,5   &   4,122121    &   340         & Ori                     \\
-161    &   -8      &   7,213711    &   ?           &       \\
-160,5  &   -9      &   6,183181    &   ?           &       \\
68      &   -8      &   8,244241    &   ?           &       \\
-76     &   6       &   -4,943996   &   ?           &       \\
-77,5   &   5       &   -3,295996   &   ?           &       \\
-17     &   8       &   4,120004    &   ?           &       \\
-18,5   &   9       &   3,296004    &   140         & Lup$^{\mathrm{b}}$      \\
-24     &   8       &   4,944004    &   140         & Lup$^{\mathrm{b}}$      \\
-1      &   8       &   4,120004    &   125         & Oph$^{\mathrm{c}}$      \\
-21,5   &   9       &   5,768004    &   140         & Lup$^{\mathrm{b}}$      \\
-15,5   &   8       &   4,120004    &   ?           &       \\
-20,5   &   8       &   2,472004    &   ?           &       \\
-29,5   &   8       &   4,944004    &   ?           &       \\
-26,5   &   9       &   4,120004    &   ?           &       \\
-30,5   &   8       &   4,944004    &   ?           &       \\
11      &   9       &   5,768004    &   ?           &       \\
-58     &   5       &   1,648004    &   ?           &       \\
-34,5   &   9       &   1,648004    &   ?           &       \\
9       &   8       &   3,296004    &   ?           &       \\
-42     &   6       &   0,000004    &   ?           &       \\
-93,5   &   -9      &   3,296004    &   380         & Vela  \\
-97     &   -8      &   9,888004    &   380         & Vela  \\
-92     &   -9      &   3,296004    &   380         & Vela  \\
-42,5   &   -5      &   -4,943996   &   ?           &       \\
-83,5   &   -9      &   2,472004    &   ?           &       \\
-72     &   -6      &   -3,295996   &   ?           &       \\
-42,5   &   -5      &   -2,471996   &   ?           &       \\
-69     &   -5      &   -2,471996   &   ?           &       \\
-85,5   &   -8      &   0,824004    &   ?           &       \\
-69     &   -6      &   -4,119996   &   ?           &       \\
-77     &   -10     &   0,000004    &   ?           &       \\
-66,5   &   -5      &   -4,119996   &   ?           &       \\
-61,5   &   -5      &   -2,471996   &   ?           &       \\
-50     &   -8      &   -0,823996   &   ?           &       \\
-16,5   &   -9      &   -2,471996   &   ?           &       \\
-55     &   -10     &   1,648004    &   ?           &       \\
\hline
\end{tabularx}
\label{tab:clumpsHI}
\begin{list}{}{}
\item[$^{\mathrm{a}}$] \citet{maddalena_1986_art}
\item[$^{\mathrm{b}}$] \citet{murphy_1986_art}
\item[$^{\mathrm{c}}$] \citet{de_geus_1990_art}
\end{list}
\end{table}

\begin{table}
\caption{CO clumps}
\begin{tabularx}{0.8\linewidth}[h]{rrrrl} \hline
 l      & b         & v$_{LSR}$     & D             & complex                 \\
 (deg)  & (deg)     & (km/s)        & (pc)          &       \\ \hline
11      &   -8      &   4,1466      &   ?           &       \\
133,5   &   9,5     &   -4,5462     &   350         & Cas$^{\mathrm{a}}$      \\
114     &   14      &   -5,8466     &   300         & Cep$^{\mathrm{a}}$      \\
-26     &   17,5    &   4,5566      &   ?           &                         \\
-170,5  &   -14     &   -0,645      &   340         & Ori                     \\
101,5   &   15      &   1,9558      &   300         & Cep$^{\mathrm{a}}$      \\
-154    &   -15     &   4,5566      &   500         & Ori$^{\mathrm{b}}$      \\
108     &   16,5    &   -4,5462     &   300         & Cep$^{\mathrm{a}}$      \\
161     &   -9      &   -4,5462     &   320         & Per$^{\mathrm{c}}$      \\
-22     &   22      &   3,2562      &   ?           &       \\
-19,5   &   8,5     &   3,2562      &   140         & Lup$^{\mathrm{d}}$      \\
125     &   12      &   0,6554      &   350         & Cas$^{\mathrm{a}}$      \\
5,5     &   19,5    &   3,2562      &   ?           &       \\
93      &   9       &   -3,2458     &   ?           &       \\
117     &   9       &   -3,2458     &   ?           &       \\
133,5   &   11      &   -3,2458     &   350         & Cas$^{\mathrm{a}}$      \\
44,5    &   8,5     &   4,5566      &   ?           &       \\
-155,5  &   -12,5   &   -0,645      &   ?           &       \\
-143,5  &   -16,5   &   3,2562      &   ?           &       \\
-154    &   -17     &   8,4578      &   500         & Ori$^{\mathrm{b}}$      \\
-150,5  &   -20     &   7,1574      &   500         & Ori$^{\mathrm{b}}$      \\
-7,5    &   16      &   1,9558      &   125         & Oph$^{\mathrm{e}}$      \\
-7      &   15      &   3,2562      &   125         & Oph$^{\mathrm{e}}$      \\
-149,5  &   -20     &   4,5566      &   ?           &       \\
-155,5  &   -14,5   &   8,4578      &   500         & Ori$^{\mathrm{b}}$      \\
-21,5   &   15,5    &   3,2562      &   ?           &       \\
158     &   -21     &   5,857       &   320         & Per$^{\mathrm{c}}$      \\
-148    &   -19,5   &   1,9558      &   ?           &       \\
-18     &   9       &   3,2562      &   140         & Lup$^{\mathrm{d}}$      \\
-5      &   20      &   1,9558      &   125         & Oph$^{\mathrm{e}}$      \\
-4      &   16      &   0,6554      &   125         & Oph$^{\mathrm{e}}$      \\
-22,5   &   16      &   4,5566      &   ?           &       \\
-5,5    &   14      &   3,2562      &   125         & Oph$^{\mathrm{e}}$      \\
-168,5  &   -11,5   &   8,4578      &   ?           &       \\
-21,5   &   14,5    &   3,2562      &   ?           &       \\
159     &   -20,5   &   5,857       &   320         & Per$^{\mathrm{c}}$      \\
-4,5    &   18,5    &   -0,645      &   ?           &       \\
160     &   -18,5   &   7,1574      &   ?           &       \\
-5,5    &   15,5    &   0,6554      &   125         & Oph$^{\mathrm{e}}$      \\
-3      &   18,5    &   0,6554      &   125         & Oph$^{\mathrm{e}}$      \\
-22     &   17      &   4,5566      &   ?           &       \\
-165,5  &   -17     &   -3,2458     &   340         & Ori      \\
-1,5    &   21      &   -0,645      &   ?           &       \\
0,5     &   9,5     &   3,2562      &   125         & Oph$^{\mathrm{e}}$      \\
112,5   &   16      &   -5,8466     &   300         & Cep$^{\mathrm{a}}$      \\
-1,5    &   18      &   0,6554      &   125         & Oph$^{\mathrm{e}}$      \\
1       &   8,5     &   1,9558      &   ?           &       \\
165     &   -9,5    &   -1,9454     &   140         & Tau$^{\mathrm{f}}$      \\
103,5   &   13,5    &   0,6554      &   300         & Cep$^{\mathrm{a}}$      \\
-154,5  &   -8,5    &   8,4578      &   500         & Ori$^{\mathrm{b}}$      \\
-143    &   -14,5   &   8,4578      &   ?           &       \\
20,5    &   9,5     &   5,857       &   250         & Aqu      \\
\hline
\end{tabularx}
\label{tab:clumpsCO}
\begin{list}{}{}
\item[$^{\mathrm{a}}$] \citet{grenier_1989_art}
\item[$^{\mathrm{b}}$] \citet{maddalena_1986_art}
\item[$^{\mathrm{c}}$] \citet{ungerechts_1987_art}
\item[$^{\mathrm{d}}$] \citet{murphy_1986_art}
\item[$^{\mathrm{e}}$] \citet{de_geus_1990_art}
\item[$^{\mathrm{f}}$] \citet{murphy_1985_art}
\end{list}
\end{table}

\begin{table}
\caption{CO clumps (continued)}
\begin{tabularx}{0.8\linewidth}[h]{rrrrl} \hline
 l      & b         & v$_{LSR}$     & D             & complex                 \\
 (deg)  & (deg)     & (km/s)        & (pc)          &       \\ \hline
132,5   &   7       &   -4,5462     &   350         & Cas$^{\mathrm{a}}$      \\
-158,5  &   -11,5   &   7,1574      &   500         & Ori$^{\mathrm{c}}$      \\
158,5   &   -9      &   -5,8466     &   320         & Per$^{\mathrm{d}}$      \\
133,5   &   9,5     &   -4,5462     &   350         & Cas       \\
114     &   14      &   -5,8466     &   300         & Cep$^{\mathrm{b}}$      \\
-26     &   17,5    &   4,5566      &   ?           &       \\
-170,5  &   -14     &   -0,645      &   340         & Ori                     \\
101,5   &   15      &   1,9558      &   300         & Cep$^{\mathrm{b}}$      \\
-154    &   -15     &   4,5566      &   500         & Ori$^{\mathrm{c}}$      \\
108     &   16,5    &   -4,5462     &   300         & Cep$^{\mathrm{b}}$      \\
161     &   -9      &   -4,5462     &   320         & Per$^{\mathrm{d}}$      \\
-22     &   22      &   3,2562      &   ?           &       \\
-19,5   &   8,5     &   3,2562      &   140         & Lup$^{\mathrm{e}}$      \\
125     &   12      &   0,6554      &   350         & Cas                     \\
5,5     &   19,5    &   3,2562      &   ?           &       \\
93      &   9       &   -3,2458     &   ?           &       \\
117     &   9       &   -3,2458     &   ?           &       \\
133,5   &   11      &   -3,2458     &   350         & Cas                     \\
44,5    &   8,5     &   4,5566      &   ?           &       \\
-155,5  &   -12,5   &   -0,645      &   ?           &       \\
-143,5  &   -16,5   &   3,2562      &   ?           &       \\
\hline
\end{tabularx}
\label{tab:clumpsCO2}
\end{table}

\end{document}